\documentclass[draftcls, onecolumn]{IEEEtran}

\ifCLASSINFOpdf
\else
\fi

\usepackage{graphicx}
\usepackage{amsmath}
\usepackage{amssymb}
\usepackage{graphics}
\usepackage{latexsym}
\usepackage[dvips]{epsfig}
\usepackage[nospace]{cite}
\usepackage{subfigure}
\usepackage{setspace}
\usepackage{tabularx}

\newtheorem{Definition}{Definition}
\newtheorem{Lemma}{Lemma}

\newtheorem{Proposition}[Lemma]{Proposition}
\newtheorem{Theorem}{Theorem}

\newtheorem{Example}{Example}
\newtheorem{Remark}{Remark}

\allowdisplaybreaks[1]

\def\Pr{{\mathrm{Pr}}}

\def\cl{{\mathrm{closure} }}

\begin{document}
%
\title{\huge Classes of Delay-Independent Multimessage Multicast Networks with Zero-Delay Nodes}
%
%
%


\author{Silas~L.~Fong,~\IEEEmembership{Member,~IEEE}
\thanks{Silas~L.~Fong is with the Department of Electrical and Computer Engineering, National University of Singapore, Singapore (e-mail: \texttt{silas\_fong@nus.edu.sg}).}
\thanks{This paper was presented in part at the International Symposium on Information Theory and Its Applications (ISITA), Melbourne, Australia, Oct., 2014.}}
\maketitle

\begin{abstract}
In a network, a node is said to \textit{incur a delay} if its encoding of each transmitted symbol involves only its received symbols obtained before the time slot in which the transmitted symbol is sent (hence the transmitted symbol sent in a time slot cannot depend on the received symbol obtained in the same time slot). A node is said to \textit{incur no delay} if its received symbol obtained in a time slot is available for encoding its transmitted symbol sent in the same time slot. Under the classical model, every node in the network incurs a delay. In this paper, we investigate the multimessage multicast network (MMN) under a generalized-delay model which allows some nodes to incur no delay. We obtain the capacity regions for three classes of MMNs with zero-delay nodes, namely the deterministic network dominated by product distribution, the MMN consisting of independent DMCs and the wireless erasure network. In addition, we show that for any MMN belonging to one of the above three classes, the set of achievable rate tuples under the generalized-delay model and under the classical model are the same, which implies that the set of achievable rate tuples for the MMN does not depend on the delay amounts incurred by the nodes in the network.
\end{abstract}


%
\IEEEpeerreviewmaketitle

\section{Introduction}\label{introduction}
%
%
%
%
\IEEEPARstart{I}{n} a multimessage multicast network (MMN), each source sends a message and each destination wants to decode all the messages. The set of source nodes and the set of destination nodes may not be disjoint. A node in the network is said to \textit{incur a delay} if its encoding of each transmitted symbol involves only its received symbols obtained before the time slot in which the transmitted symbol is sent. In contrast, a node is said to \textit{incur no delay} if its received symbol obtained in a time slot is available for encoding its transmitted symbol sent in the same time slot. Similarly, the network is said to \textit{contain zero-delay nodes} if there exists a node that incurs zero delay on another node; the network is said to \textit{contain no zero-delay node} if every node incurs a delay on all the nodes. In \cite{fongYeung15}, the \textit{capacity region} of the MMN with zero-delay nodes is defined to be the set of rate tuples achievable by all \textit{feasible} schemes that do not include deadlock loops, and the \textit{positive-delay region} is defined to be the set of rate tuples achievable by all classical schemes under the constraint that each node incurs a delay (and hence deadlock loops are automatically excluded). By this definition, the positive-delay region is always a subset of the capacity region.

It is easy to construct a network with zero-delay nodes whose capacity region is strictly larger than the positive-delay region. One such network is the binary symmetric channel with correlated feedback (BSC-CF) considered in \cite[Section VII]{fongYeung15}, which will be introduced in the next section.

\subsection{A Motivating Example} \label{sectionMotivating}
\begin{figure}[!t]
 \centering
  \includegraphics[width=1.8 in, height=1.1 in, bb = 203 258 468 425, angle=0, clip=true]{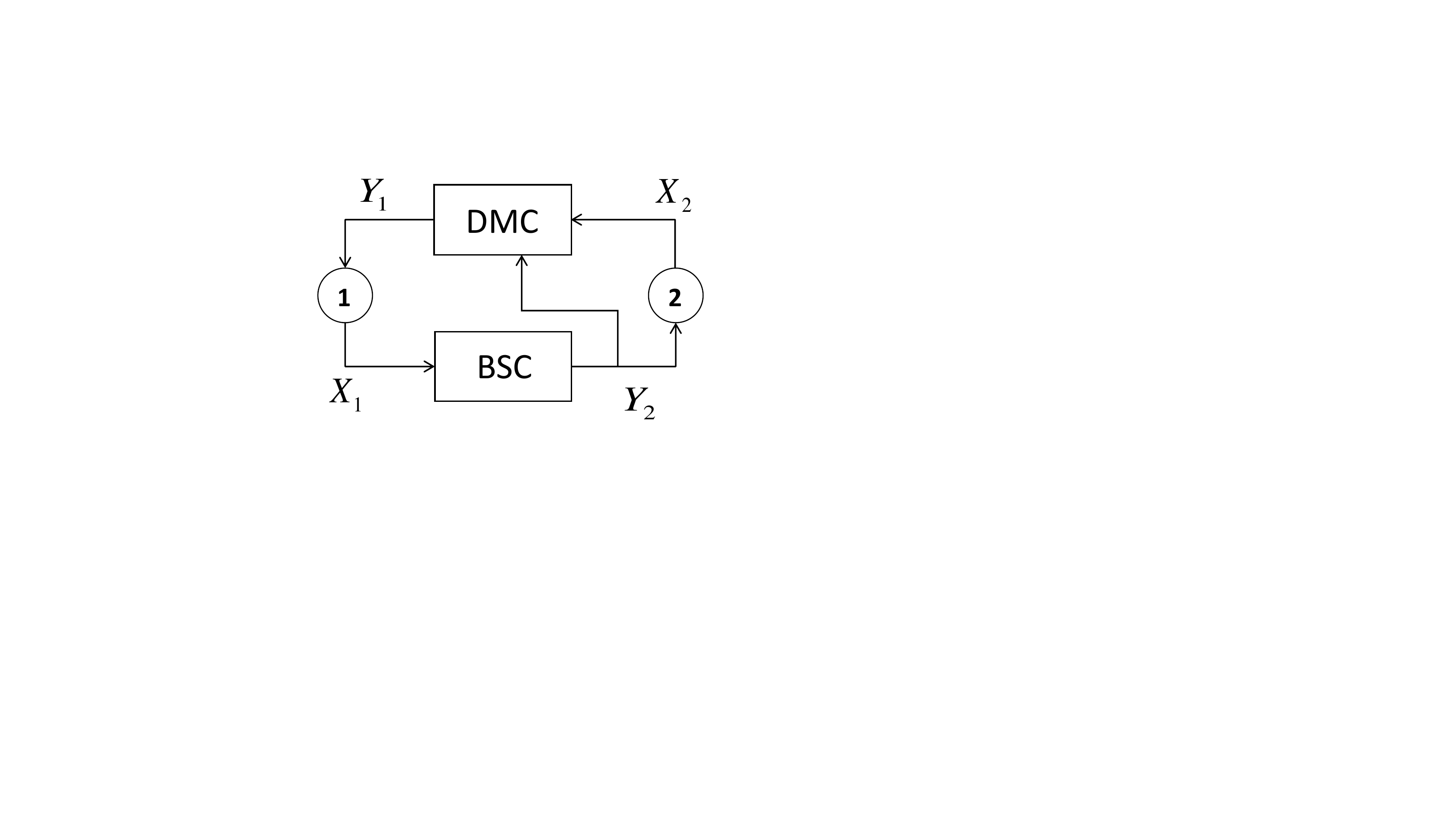}
\caption{BSC-DMCF / BSC-CF}  \label{BSCFB}
\end{figure}

Consider a network that consists of two nodes denoted by $1$ and $2$ respectively. Node~1 and node~2 want to transmit a message to each other. This is a two-way channel \cite{database:bib17}. Since we can assume without loss of generality that both nodes want to decode both messages, this network can be regarded as a MMN.
In each time slot, node~1 and node~2 transmit $X_1$ and $X_2$ respectively, and they receive $Y_1$ and $Y_2$ respectively. All the input and output alphabets are binary, and the channel that carries $X_1$ to node~2 is a binary symmetric channel (BSC) while the channel that carries $X_2$ to node~1 is a discrete memoryless channel (DMC) whose output may depend on the output of the BSC. In this network, node~1 incurs zero delay on node~2, i.e., node~$2$ can receive $Y_2$ before encoding and transmitting $X_2$. We call this network the \textit{BSC with DMC feedback} (BSC-DMCF), which is illustrated in Figure~\ref{BSCFB}.
The BSC-CF is a special case of the BSC-DMCF when $Y_1=X_2+Y_2$  \cite[Section VII]{fongYeung15}, where~$+$ denotes the XOR operation. It is shown in \cite[Section IX]{fongYeung15} that the capacity region of the BSC-CF is strictly larger than the positive-delay region (recall that the positive-delay region is obtained under the assumption that the network contains no zero-delay node while the capacity region of the BSC-CF is achieved when node~1 incurs zero delay on node~2).
Other MMNs whose capacity regions are strictly larger than their positive-delay regions include the \textit{relay-without-delay channel} studied by El~Gamal \textit{et al}.\ \cite{AbbasRelayNetwork} and the \textit{causal relay network} studied by Baik and Chung \cite{causalRelayNetwork}. In other words, for some MMNs with zero-delay nodes, their capacity regions can be strictly larger than their positive-delay regions, which motivates us to classify the set of MMNs with zero-delay nodes into the following two categories:
\begin{description}
\item[(i)]\textit{Delay-independent MMNs} whose capacity regions coincide with their positive-delay regions.
\item[(ii)]   \textit{Delay-dependent MMNs} whose capacity regions are strictly larger than their positive-delay regions.
\end{description}
For each MMN in Category~(i), the set of achievable rate tuples does not depend on the delay amounts incurred by the nodes in the network. On the other hand, for each MMN in Category~(ii), the set of achievable rate tuples shrinks if we impose the additional constraint that each node incurs a positive delay. It is important to decide which category a given MMN belongs to because the category of the MMN affects how the delays should be handled and how the transmissions in the network should be synchronized to achieve optimal performance.
\subsection{Main Contribution}
The main contributions of this work are identification of three classes of delay-independent MMNs and complete characterizations of their capacity regions. The first class is called the \textit{deterministic MMN dominated by product distribution}. Being a subclass of MMNs consisting of deterministic channels, the deterministic MMN dominated by product distribution is a generalization of the deterministic relay network with no interference in \cite{multicastCapacityRelayNetworks} and the finite-field linear deterministic network in \cite{AvestimehrDeterministic,linearFiniteField09}. The second class is the \textit{MMN consisting of independent DMCs} \cite{networkEquivalencePartI}. The third class is the \textit{wireless erasure network} \cite{dana06}. We successfully evaluate the capacity regions for the above classes of MMNs with zero-delay nodes and show that their capacity regions coincide with their positive-delay regions, which implies that the above classes of MMNs belong to the category of delay-independent MMNs. A natural consequence of our result is that for any MMN belonging to one of the above three classes, using different methods for handling delays and synchronization in the network does not affect the capacity region.

Given a MMN with zero-delay nodes belonging to one of the above three classes, in order to show its delay-independence, we first evaluate an achievable rate region for the MMN by invoking the noisy network coding (NNC) inner bound \cite[Theorem~1]{noisyNetworkCoding} (which was also discovered in \cite{NNCv2}). The achievable rate region is contained in the positive-delay region because the NNC inner bound was proved in \cite{noisyNetworkCoding} for classical MMNs. Then, we evaluate an outer bound on the capacity region of the MMN with zero-delay nodes by simplifying the cut-set outer bound in \cite[Theorem~1]{fongYeung15} and show that the cut-set outer bound coincides with the NNC inner bound (which is within the positive-delay region), implying that the MMN is delay-independent.

This work should not be confused with the work by Effros \cite{EffrosIndependentDelay}, which shows that under the classical model which assumes a positive delay at every node, the set of achievable rate tuples for any MMN does not depend on the amount of positive delay incurred by each node. Here, we prove a different result for the above classes of MMNs with zero-delay nodes that their capacity regions and positive-delay regions are the same. Our result is meaningful given the fact that for some MMNs with zero-delay nodes, their capacity regions are strictly larger than their positive-delay regions (see Section~\ref{sectionMotivating}).

\subsection{Paper Outline}
This paper is organized as follows. Section~\ref{notation} presents the notation used in this paper. Section~\ref{sectionDefinition} presents the formulation of the MMN with zero-delay nodes. Section~\ref{sectionInnerOuterBound} recapitulates the NNC inner bound and the cut-set outer bound for the capacity region of the MMN with zero-delay nodes. In Section~\ref{sectionClassesDelayIndependent}, we use the two bounds obtained in Section~\ref{sectionInnerOuterBound} to identify the three classes of delay-independent MMNs -- the deterministic MMN dominated by product distributions, the MMN consisting of independent DMCs and the wireless erasure network, whose problem formulations and proofs for delay-independence are contained in Section~\ref{sectionDeterministicNetwork}, Section~\ref{sectionDM-MMNconsistingOfDMCs} and Section~\ref{sectionErasureNetworks} respectively. Concluding remarks are given in Section~\ref{conclusion}.

\section{Notation}\label{notation}
We use $\Pr\{\mathcal{E}\}$ to represent the probability of an
event~$\mathcal{E}$, and use $\mathbf{1}\{\mathcal{E}\}$ to denote the characteristic function of $\mathcal{E}$. We use a capital letter~$X$ to denote a random variable with alphabet $\mathcal{X}$, and use the small letter $x$ to denote the realization of~$X$.
We use $X^n$ to denote a random tuple $(X_1,  X_2,  \ldots,  X_n)$, where the components $X_k$ have the same alphabet~$\mathcal{X}$.
 We let $p_X$ and $p_{Y|X}$ denote the probability mass distribution of $X$ and the conditional probability mass distribution of $Y$ given $X$ respectively for any discrete random variables~$X$ and~$Y$. We let $p_X(x)\triangleq\Pr\{X=x\}$ and $p_{Y|X}(y|x)\triangleq\Pr\{Y=y|X=x\}$ be the evaluations of $p_X$ and $p_{Y|X}$ respectively at $X=x$ and $Y=y$. We let $p_Xp_{Y|X}$ denote the joint distribution of $(X,Y)$, i.e., $p_Xp_{Y|X}(x,y)=p_X(x)p_{Y|X}(y|x)$ for all $x$ and $y$. If $X$ and $Y$ are independent, their joint distribution is simply $p_X p_Y$. We will take all logarithms to base 2.
For any discrete random variable $(X,Y,Z)$ distributed according to $p_{X,Y,Z}$, we let $H_{p_{X,Z}}(X|Z)$ and $I_{p_{X,Y,Z}}(X;Y|Z)$ be the entropy of $X$ given $Z$ and mutual information between $X$ and $Y$ given $Z$  respectively. For simplicity, we drop the subscript of a notation if there is no ambiguity.
  If $X$, $Y$ and $Z$ are distributed according to $p_{X,Y,Z}$ and they form a Markov chain, we write
$(X\rightarrow Y\rightarrow Z)_{p_{X,Y,Z}}$ or more simply, $(X\rightarrow Y\rightarrow Z)_p$. The sets of natural and real numbers are denoted by $\mathbb{N}$ and $\mathbb{R}$ respectively. The closure of a set $S$ is denoted by $\cl(S)$


\section{Discrete Memoryless Multimessage Multicast Network with Zero-Delay Nodes} \label{sectionDefinition}
We consider a multimessage multicast network (MMN) that consists of $N$ nodes. Let
\[
\mathcal{I}\triangleq \{1, 2, \ldots, N\}
\]
be the index set of the nodes, and let $\mathcal{V}\subseteq \mathcal{I}$ and $\mathcal{D}\subseteq \mathcal{I}$ be the sets of sources and destinations respectively. We call $(\mathcal{V}, \mathcal{D})$ the \textit{multicast demand} on the network. The sources in $\mathcal{V}$ transmit information to the destinations in $\mathcal{D}$ in $n$ time slots (channel uses) as follows.
Each node~$i\in \mathcal{V}$ transmits message
\[
W_{i}\in \{1, 2, \ldots, M_i\}
 \]
and each node $j \in\mathcal{D}$ wants to decode all the messages $\{W_{i}: i\in \mathcal{V}\}$. We assume that each message $W_{i}$ is uniformly distributed over $\{1, 2, \ldots, M_i\}$ and all the messages are independent. For each $k\in \{1, 2, \ldots, n\}$ and each $i\in \mathcal{I}$, node~$i$ transmits $X_{i,k} \in \mathcal{X}_i$ and receives $Y_{i,k} \in \mathcal{Y}_i$ in the $k^{\text{th}}$ time slot where $\mathcal{X}_i$ and $\mathcal{Y}_i$ are some alphabets that depend on~$i$.
After~$n$ time slots, node~$j$ declares~$\hat W_{i,j}$ to be the
transmitted~$W_{i}$ based on $(W_{j},Y_j^n)$ for each $(i, j)\in \mathcal{V}\times \mathcal{D}$.

To simplify notation, we use the following conventions for each $T\subseteq \mathcal{I}$:
For any random tuple \[(X_{1}, X_{2}, \ldots, X_{N}) \in \mathcal{X}_1\times \mathcal{X}_2 \times \ldots \times \mathcal{X}_N,\] we let
\[
 X_T \triangleq (X_{i} : i\in T)\] be a subtuple of $(X_{1}, X_{2}, \ldots, X_{N})$.
Similarly, for any $k\in \{1, 2, \ldots, n\}$ and any random tuple
\[
(X_{1,k}, X_{2,k}, \ldots, X_{N, k}) \in \mathcal{X}_1\times \mathcal{X}_2 \times \ldots \times \mathcal{X}_N,
\]
 we let
\[
X_{T,k}\triangleq(X_{i,k} : i\in T)
 \]
 be a subtuple of $(X_{1,k}, X_{2,k}, \ldots, X_{N, k})$.
For any $N^2$-dimensional random tuple $(\hat W_{1,1}, \hat W_{1,2}, \ldots, \hat W_{N,N})$, we let
\[
\hat W_{T\times T^c}\triangleq(\hat W_{i,j} : (i,j)\in T\times T^c)
\]
 be a subtuple of $(\hat W_{1,1}, \hat W_{1,2}, \ldots, \hat W_{N,N})$.

We follow the formulation of the discrete memoryless network with zero-delay nodes in \cite{fongYeung15}, which includes the following six definitions. The definitions are given here for completeness, and the detailed motivations behind them can be found in \cite{fongYeung15}.
\medskip
\begin{Definition}\label{defOrderedPartition}
An $\alpha$-dimensional tuple $(\mathcal{S}_1, \mathcal{S}_2, \ldots \mathcal{S}_\alpha)$ consisting of subsets of $\mathcal{I}$ is called an \textit{$\alpha$-partition of $\mathcal{I}$} if $\cup_{h=1}^\alpha \mathcal{S}_h = \mathcal{I}$ and $\mathcal{S}_i\cap \mathcal{S}_j = \emptyset$ for all $i\ne j$.
\end{Definition}
\medskip

For any $(\mathcal{S}_1, \mathcal{S}_2, \ldots \mathcal{S}_\alpha)$ which is an $\alpha$-partition of $\mathcal{I}$, we let
\[
\mathcal{S}^h \triangleq \cup_{i=1}^h \mathcal{S}_i
\]
for each $h\in\{1, 2, \ldots, \alpha\}$ to facilitate discussion.
 \medskip
\begin{Definition} \label{defDiscreteNetwork}
The discrete network consists of $N$ finite input sets
$\mathcal{X}_1, \mathcal{X}_2, \ldots, \mathcal{X}_N$, $N$ finite output sets $\mathcal{Y}_1, \mathcal{Y}_2, \ldots, \mathcal{Y}_N$ and $\alpha$ channels characterized by conditional distributions $q_{Y_{\mathcal{G}_1}|X_{\mathcal{S}^1}}^{(1)}$, $q_{Y_{\mathcal{G}_2}|X_{\mathcal{S}^2}, Y_{\mathcal{G}^1}}^{(2)},\ldots,q_{Y_{\mathcal{G}_\alpha}|X_{\mathcal{S}^\alpha},Y_{\mathcal{G}^{\alpha-1}}}^{(\alpha)}$, where
\[
\boldsymbol{\mathcal{S}} \triangleq (\mathcal{S}_1, \mathcal{S}_2, \ldots \mathcal{S}_\alpha)
 \]and
 \[
 \boldsymbol{\mathcal{G}} \triangleq  (\mathcal{G}_1, \mathcal{G}_2, \ldots \mathcal{G}_\alpha)
  \]
  are two $\alpha$-dimensional partitions of $\mathcal{I}$. We call $\boldsymbol{\mathcal{S}}$ and $\boldsymbol{\mathcal{G}}$ the \textit{input partition} and the \textit{output partition} of the network respectively. The discrete network is denoted by $(\mathcal{X}_\mathcal{I}, \mathcal{Y}_\mathcal{I}, \alpha, \boldsymbol{\mathcal{S}}, \boldsymbol{\mathcal{G}}, \boldsymbol q)$ where
  \[
  \boldsymbol q \triangleq (q^{(1)}, q^{(2)}, \ldots, q^{(\alpha)}).
  \]
\end{Definition}
\medskip
\begin{Definition} \label{defDelayProfile}
A delay profile is an $N$-dimensional tuple $(b_1, b_2, \ldots , b_N)$ where $b_i \in \{0, 1\}$ for each $i\in \mathcal{I}$. The delay profile is said to be \textit{positive} if its elements are all 1.
\end{Definition}
\medskip

When we formally define a code on the discrete network later, a delay profile $B=(b_1, b_2, \ldots , b_N)$ will be associated with the code and $b_i$ represents the amount of delay incurred by node~$i$ for the code. Under the classical model, $B$ can only be positive, meaning that the amount of delay incurred by each node is positive. In contrast, under our generalized-delay model some elements of $B$ can take $0$ as long as deadlock loops do not occur. Therefore our model is a generalization of the classical model. The essence of the following definition is to characterize delay profiles which will not cause deadlock loops for the transmissions in the network.
\medskip
\begin{Definition} \label{defFeasible}
Let $(\mathcal{X}_\mathcal{I}, \mathcal{Y}_\mathcal{I}, \alpha, \boldsymbol{\mathcal{S}}, \boldsymbol{\mathcal{G}}, \boldsymbol q)$ be a discrete network. For each $i\in \mathcal{I}$, let $h_i$ and $m_i$ be the two unique integers such that $i\in \mathcal{S}_{h_i}$ and $i\in\mathcal{G}_{m_i}$. Then, a delay profile $(b_1, b_2, \ldots, b_N)$ is said to be \textit{feasible for the network} if the following holds for each $i\in \mathcal{I}$: If $b_i=0$, then $h_i > m_i$.
\end{Definition}
\medskip
Under the classical model, Definition~\ref{defFeasible} is trivial because any delay profile is positive and hence always feasible for the network. We are ready to define codes that use the network $n$ times as follows.
\medskip
\begin{Definition} \label{defCode}
Let $B\triangleq (b_1, b_2, \ldots , b_N)$ be a delay profile feasible for $(\mathcal{X}_\mathcal{I}, \mathcal{Y}_\mathcal{I}, \alpha, \boldsymbol{\mathcal{S}}, \boldsymbol{\mathcal{G}}, \boldsymbol q)$, and let $(\mathcal{V}, \mathcal{D})$ be the multicast demand on the network. A $(B, n, M_{\mathcal{I}})$-code, where $M_\mathcal{I}\triangleq (M_1, M_2, \ldots, M_N)$ denotes the tuple of message alphabets, for $n$ uses of the network consists of the following:
\begin{enumerate}
\item A message set
\[
\mathcal{W}_{i}\triangleq \{1, 2, \ldots, M_i\}
\]
 at node~$i$ for each $i\in \mathcal{I}$, where $M_i=1$ for each $i\in \mathcal{V}^c$. Message $W_i$ is uniformly distributed on $\mathcal{W}_i$.

\item An encoding function
$
f_{i,k} : \mathcal{W}_i \times \mathcal{Y}_i^{k-b_i} \rightarrow \mathcal{X}_i
$
 for each $i\in \mathcal{I}$ and each $k\in\{1, 2, \ldots, n\}$, where $f_{i,k}$ is the encoding function at node~$i$ in the
$k^{\text{th}}$ time slot such that
$
X_{i,k}=f_{i,k} (W_{i},
Y_i^{k-b_i})$.

\item A decoding function
$
g_{i,j} : \mathcal{W}_{j}  \times
\mathcal{Y}_j^{n} \rightarrow \mathcal{W}_{i}
$
 for each $(i, j) \in \mathcal{V}\times \mathcal{D}$, where $g_{i,j}$ is the decoding function for $W_{i}$ at node~$j$ such that
\[
 \hat W_{i, j} \triangleq g_{i,j}(W_{j}, Y_j^{n}).
 \]
\end{enumerate}
\end{Definition}
\medskip

Given a $(B, n, M_{\mathcal{I}})$-code, it follows from Definition~\ref{defCode} that for each $i\in\mathcal{I}$, node~$i$ incurs a delay if $b_i>0$, where $b_i$ is the amount of delay incurred by node~$i$. If $b_i=0$, node~$i$ incurs no delay, i.e., for each $k\in \{1, 2, \ldots, n\}$, node~$i$ needs to receive $Y_{i,k}$ before encoding $X_{i,k}$. The feasibility condition of $B$ in Definition~\ref{defFeasible} ensures that the operations of any $(B, n, M_{\mathcal{I}})$-code are well-defined for the subsequently defined discrete memoryless network; the associated coding scheme is described after the network is defined.
\medskip
\begin{Definition}\label{defDiscreteMemoryless}
A discrete network $(\mathcal{X}_\mathcal{I}, \mathcal{Y}_\mathcal{I}, \alpha, \boldsymbol{\mathcal{S}}, \boldsymbol{\mathcal{G}}, \boldsymbol q)$ with multicast demand $(\mathcal{V}, \mathcal{D})$, when used multiple times, is called a \textit{discrete memoryless multimessage multicast network (DM-MMN)} if the following holds for any $(B, n, M_{\mathcal{I}})$-code:

Let $U^{k-1}\triangleq (W_{\mathcal{I}}, X_{\mathcal{I}}^{k-1}, Y_{\mathcal{I}}^{k-1})$ be the collection of random variables that are generated before the $k^{\text{th}}$ time slot. Then, for each $k\in\{1, 2, \ldots, n\}$ and each $h\in \{1, 2, \ldots, \alpha\}$,
\begin{align}
& \Pr\{U^{k-1} = u^{k-1}, X_{\mathcal{S}^h,k} =x_{\mathcal{S}^h,k}, Y_{\mathcal{G}^{h},k}=y_{\mathcal{G}^{h},k} \}
 \notag\\
 & =  \Pr\{U^{k-1} = u^{k-1}, X_{\mathcal{S}^h,k} =x_{\mathcal{S}^h,k}, Y_{\mathcal{G}^{h-1},k}=y_{\mathcal{G}^{h-1},k} \} q_{Y_{\mathcal{G}_h}|X_{\mathcal{S}^h}, Y_{\mathcal{G}^{h-1}}}^{(h)}(y_{\mathcal{G}_{h},k}|x_{\mathcal{S}^h,k}, y_{\mathcal{G}^{h-1},k}) \label{memorylessStatement}
\end{align}
for all $u^{k-1}\in \mathcal{U}^{k-1}$, $x_{\mathcal{S}^h,k}\in \mathcal{X}_{\mathcal{S}^h}$ and $y_{\mathcal{G}^h,k}\in \mathcal{Y}_{\mathcal{G}^h}$.
\end{Definition}
\medskip

Following the notation in Definition~\ref{defDiscreteMemoryless}, consider any $(B, n, M_{\mathcal{I}})$-code on the DM-MMN. In the $k^{\text{th}}$ time slot, $X_{\mathcal{I},k}$ and $Y_{\mathcal{I},k}$ are generated in the order
\begin{equation}
X_{\mathcal{S}_1,k}, Y_{\mathcal{G}_1,k}, X_{\mathcal{S}_2,k}, Y_{\mathcal{G}_2,k}, \ldots, X_{\mathcal{S}_\alpha,k}, Y_{\mathcal{G}_\alpha,k} \label{orderExplanation}
\end{equation}
by transmitting on the channels in this order  $q^{(1)},q^{(2)}, \ldots, q^{(\alpha)}$ using the $(B, n, M_{\mathcal{I}})$-code (as prescribed in Definition~\ref{defCode}). Specifically, $X_{\mathcal{S}^h, k}$, $Y_{\mathcal{G}^{h-1},k}$ and channel $q^{(h)}$ together define $Y_{\mathcal{G}_h,k}$ for each $h\in\{1, 2, \ldots, \alpha\}$. It is shown in \cite[Section IV]{fongYeung15} that the encoding of $X_{\mathcal{S}_h,k}$ before the transmission on $q^{(h)}$ and the generation of $Y_{\mathcal{G}_h,k}$ after the transmission on $q^{(h)}$ for each $h\in\{1, 2, \ldots, \alpha\}$ are well-defined.

After defining the DM-MMN with zero-delay nodes in the above six definitions, we are now ready to formally define the \textit{capacity region} and the \textit{positive-delay region} through the following three intuitive definitions.
\smallskip
\begin{Definition} \label{defErrorProbability}
For a $(B, n, M_{\mathcal{I}})$-code on the DM-MMN, the average probability of decoding error $P_{\text{err}}^{n}$ is defined as
\[
P_{\text{err}}^{n} \triangleq \Pr\bigg\{\bigcup_{(i,j)\in \mathcal{V}\times \mathcal{D}} \big\{\hat{W}_{i,j} \ne W_{i } \big\}\bigg\}.
\]
\end{Definition}
\medskip
%
\begin{Definition} \label{defAchievableRate}
Let $B$ be a feasible delay profile for the network. A rate tuple $(R_{1}, R_{2}, \ldots, R_{N})$, denoted by $R_{\mathcal{I}}$, is \textit{$B$-achievable} for the DM-MMN if there exists a sequence of $(B, n, M_{\mathcal{I}})$-codes such that
\[
 \lim\limits_{n\rightarrow \infty} \frac{\log M_{i}}{n} \ge R_{i}
\]
for each $i\in\mathcal{I}$ and
\[
 \lim\limits_{n\rightarrow \infty} P_{\text{err}}^{n} = 0.
\]
\end{Definition}
\medskip

\begin{Definition}\label{defCapacityRegion}
The \textit{$B$-capacity region}, denoted by $\mathcal{C}_B$, of the DM-MMN is the set consisting of every $B$-achievable rate tuple $R_{\mathcal{I}}$ with $R_{i}=0$ for all $i\in \mathcal{V}^c$. The \textit{capacity region} $\mathcal{C}$ is defined as
\[
\mathcal{C}\triangleq \bigcup_{B: B\text{ is feasible}}\mathcal{C}_B
\]
and the \textit{positive-delay region} $\mathcal{C}_+$ is defined as
\[
\mathcal{C}_+ \triangleq \bigcup_{B: B\text{ is positive}}\mathcal{C}_B.
\]
If $\mathcal{C} = \mathcal{C}_+$, the DM-MMN is said to be \textit{delay-independent}. If $\mathcal{C} \supsetneq \mathcal{C}_+$, the DM-MMN is said to be \textit{delay-dependent}.
\end{Definition}
\medskip
Roughly speaking, the capacity region is the set of rate tuples which are achievable by codes that do not incur a deadlock loop, and the positive-delay region is the set of rate tuples which are achievable by codes under the constraint that every node incurs a delay.
Definitions~\ref{defDelayProfile},~\ref{defFeasible} and~\ref{defCapacityRegion} imply that $\mathcal{C} \supseteq \mathcal{C}_+$, which implies that each DM-MMN is either delay-independent (i.e., $\mathcal{C} = \mathcal{C}_+$) or delay-dependent (i.e., $\mathcal{C} \supsetneq \mathcal{C}_+$).

\section{Inner and Outer Bounds for the Capacity Region}\label{sectionInnerOuterBound}
We start this section by stating an achievability result for classical DM-MMNs in the following theorem, which is a specialization of the main result of {\em noisy network coding (NNC)} inner bound by Lim, Kim, El Gamal and Chung~\cite{noisyNetworkCoding} (the NNC inner bound was also discovered by Yassaee and Aref~\cite{NNCv2}).

\begin{Theorem} \label{theoremNNC}
 Let $(\mathcal{X}_\mathcal{I}, \mathcal{Y}_\mathcal{I}, \alpha, \boldsymbol{\mathcal{S}}, \boldsymbol{\mathcal{G}}, \boldsymbol q)$ be a DM-MMN, and let
\begin{align}
\mathcal{R}_{\text{in}} & \triangleq   \bigcup_{\substack{p_{X_\mathcal{I}, Y_\mathcal{I}}: p_{X_\mathcal{I}, Y_\mathcal{I}}= \\ (\prod_{i=1}^N p_{X_i})(\prod_{h=1}^\alpha q_{Y_{\mathcal{G}_h}|X_\mathcal{S}^{h}, Y_\mathcal{G}^{h-1}}^{(h)})}} \bigcap_{T\subseteq \mathcal{I}: T^c \cap \mathcal{D} \ne \emptyset } \notag\\*
& \hspace{0.5 in}\left\{ R_\mathcal{I}\left| \: \parbox[c]{3.6 in}{$ \sum_{ i\in T}  R_{i}
 \le  I_{p_{X_\mathcal{I},Y_\mathcal{I}}}(X_T; Y_{T^c}|X_{T^c})-H_{p_{X_\mathcal{I},Y_\mathcal{I}}}(Y_T|X_\mathcal{I}, Y_{T^c}), \\
 R_i=0 \text{ for all }i\in\mathcal{V}^c$} \right.\right\}. \label{Rin}
\end{align}
Then,
\begin{equation}
\mathcal{R}_{\text{in}} \subseteq \mathcal{C}_+. \label{theoremSt1theoremNNC}
\end{equation}
\end{Theorem}
\begin{IEEEproof}
For every (classical) $((1, 1, \ldots, 1), n, M_\mathcal{I})$-code, the MMN $(\mathcal{X}_\mathcal{I}, \mathcal{Y}_\mathcal{I}, \alpha, \boldsymbol{\mathcal{S}}, \boldsymbol{\mathcal{G}}, \boldsymbol q)$ is equivalent to the MMN $(\mathcal{X}_\mathcal{I}, \mathcal{Y}_\mathcal{I}, 1, \mathcal{I}, \mathcal{I} , \prod_{h=1}^\alpha q_{Y_{\mathcal{G}_h}|X_\mathcal{S}^{h}, Y_\mathcal{G}^{h-1}}^{(h)})$ by Theorem~3 in \cite{fongYeung15}. The intuition behind the above equivalence can be reasoned as follows: If every node incurs a delay, then the outputs of the $\alpha$ channels in $\boldsymbol q$ will be independent given their inputs, and hence the relationship between the inputs and outputs of the network can be characterized simply by a product of the $\alpha$ channels, which is $\prod_{h=1}^\alpha q_{Y_{\mathcal{G}_h}|X_\mathcal{S}^{h}, Y_\mathcal{G}^{h-1}}^{(h)}$.

On the other hand, $\mathcal{R}_{\text{in}}$ is a specialization of the NNC inner bound in \cite[Theorem 1]{noisyNetworkCoding} by taking $\hat Y=Y$ for the MMN $(\mathcal{X}_\mathcal{I}, \mathcal{Y}_\mathcal{I}, 1, \mathcal{I}, \mathcal{I} , \prod_{h=1}^\alpha q_{Y_{\mathcal{G}_h}|X_\mathcal{S}^{h}, Y_\mathcal{G}^{h-1}}^{(h)})$. Since the NNC inner bound was developed under the classical model where each node incurs a delay, any rate tuple in $\mathcal{R}_{\text{in}}$ is achievable by some sequence of $((1, 1, \ldots, 1), n, M_\mathcal{I})$-codes, which then implies \eqref{theoremSt1theoremNNC}.
\end{IEEEproof}
\medskip

Following similar procedures for proving Theorem~1 in \cite{fongYeung15}, we can prove an outer bound on $\mathcal{C}$ stated in the following theorem. Since networks with zero-delay nodes can be viewed as networks with in-block memory formulated by Kramer \cite[Section VII-D]{kramer14}, the following theorem can be seen as the multicast version of Theorem~1 in \cite{kramer14}.
\smallskip
\begin{Theorem} \label{thmCapacityRegionMulticast}
 Let $(\mathcal{X}_\mathcal{I}, \mathcal{Y}_\mathcal{I}, \alpha, \boldsymbol{\mathcal{S}}, \boldsymbol{\mathcal{G}}, \boldsymbol q)$ be a DM-MMN, and let
 \begin{align}
\mathcal{R}_{\text{out}}& \triangleq   \bigcup_{\substack{p_{X_{\mathcal{I}}, Y_{\mathcal{I}}}:p_{X_{\mathcal{I}}, Y_{\mathcal{I}}}=\\ \prod_{h=1}^\alpha (p_{X_{\mathcal{S}_h}|X_{\mathcal{S}^{h-1}}, Y_{\mathcal{G}^{h-1}}} q^{(h)}_{Y_{\mathcal{G}_h} | X_{\mathcal{S}^h},  Y_{\mathcal{G}^{h-1}}}) }} \bigcap_{T\subseteq \mathcal{I}: T^c \cap \mathcal{D} \ne \emptyset } \notag\\*
& \quad \qquad \left\{ R_\mathcal{I}\left| \: \parbox[c]{4.25 in}{$ \sum_{ i\in T}  R_{i}
 \le \sum_{h=1}^\alpha  I_{p_{X_{\mathcal{I}}, Y_{\mathcal{I}}}}(X_{T\cap \mathcal{S}^h}, Y_{T\cap \mathcal{G}^{h-1}} ;  Y_{T^c\cap \mathcal{G}_h}| X_{T^c \cap \mathcal{S}^h},  Y_{T^c\cap \mathcal{G}^{h-1}}),\\
 R_i=0 \text{ for all }i\in\mathcal{V}^c$} \right.\right\}. \label{Rout}
\end{align}
Then,
\[\mathcal{C}\subseteq\mathcal{R}_{\text{out}}.
\]
\end{Theorem}
\begin{IEEEproof}
Let $R_{\mathcal{I}}$ be an achievable rate tuple for the DM-MMN denoted by $(\mathcal{X}_\mathcal{I}, \mathcal{Y}_\mathcal{I}, \alpha, \boldsymbol{\mathcal{S}}, \boldsymbol{\mathcal{G}}, \boldsymbol q)$. By Definitions~\ref{defAchievableRate} and~\ref{defCapacityRegion}, there exists a sequence of $(B, n, M_{\mathcal{I}})$-codes on the DM-MMN such that
\begin{equation}
 \lim_{n\rightarrow \infty} \frac{\log M_{i}}{n} \ge R_{i} \label{thmTempEq1}
\end{equation}
for each $i\in \mathcal{I}$ and
\begin{equation}
 \lim_{n\rightarrow \infty} P_{\text{err}}^n = 0. \label{thmTempEq2}
\end{equation}

Fix any $T\subseteq \mathcal{I}$ such that $T^c\cap \mathcal{D}\ne \emptyset$, and let $d$ denote a node in $T^c\cap \mathcal{D}$.
For each $(B, n, M_{\mathcal{I}})$-code, since the $N$ messages $W_{1}, W_{2}, \ldots, W_{N}$ are independent, we have
\begin{align}
\sum_{i\in T} \log M_{i}
& = H(W_{T}|W_{T^c})\notag \\
&= I(W_{T}; Y_{T^c}^n|W_{T^c}) +
H(W_{T}|Y_{T^c}^n,W_{T^c}) \notag\\
&\le I(W_{T}; Y_{T^c}^n|W_{T^c}) + H(W_{T}|Y_d^n, W_{d})\notag \\
 &\le   I(W_{T};  Y_{T^c}^n|W_{T^c}) + 1+ P_{\text{err}}^n \sum_{i\in T} \log M_{i},
\label{cutseteqnSet1}
\end{align}
where the last inequality follows from Fano's inequality (cf.\ Definition~\ref{defErrorProbability}). Following similar procedures for proving Theorem 1 in \cite{fongYeung15}, we can show by using \eqref{thmTempEq1}, \eqref{thmTempEq2} and \eqref{cutseteqnSet1} that there exists a joint distribution $p_{X_\mathcal{I}, Y_\mathcal{I}}$ which depends on the sequence of $(B, n, M_{\mathcal{I}})$-codes but not on $T$ such that
\[
  p_{X_{\mathcal{I}}, Y_{\mathcal{I}}} =
\prod_{h=1}^\alpha (p_{X_{\mathcal{S}_h}|X_{\mathcal{S}^{h-1}}, Y_{\mathcal{G}^{h-1}}}
q^{(h)}_{Y_{\mathcal{G}_h} | X_{\mathcal{S}^h},  Y_{\mathcal{G}^{h-1}}})
\]
and
\begin{equation}
\sum_{i\in T} R_i\le  \sum_{h=1}^\alpha  I_{p_{X_{\mathcal{I}}, Y_{\mathcal{I}}}}(X_{T\cap \mathcal{S}^h}, Y_{T\cap \mathcal{G}^{h-1}} ;  Y_{T^c\cap \mathcal{G}_h}| X_{T^c \cap \mathcal{S}^h},  Y_{T^c\cap \mathcal{G}^{h-1}}). \label{stInTheorem}
\end{equation}
Since $p_{X_\mathcal{I}, Y_\mathcal{I}}$ depends on only the sequence of $(B, n, M_{\mathcal{I}})$-codes but not on $T$, \eqref{stInTheorem} holds for all $T\subseteq \mathcal{I}$ such that $T^c\cap \mathcal{D}\ne \emptyset$. This completes the proof.
\end{IEEEproof}

\section{Classes of Delay-Independent MMNs} \label{sectionClassesDelayIndependent}
In this section, we will use our inner and outer bounds developed in the previous section to calculate the capacity regions for some classes of MMNs with zero-delay nodes and then show that the MMNs are delay-independent, i.e., $\mathcal{C}=\mathcal{C}_+$. In the process of calculating their capacity regions, we will use the following proposition extensively to characterize an important property of Markov chains.
\begin{Proposition} \label{propositionMCsimplification}
Suppose there exist two probability distributions $r_{X,Y}$ and $q_{Z|Y}$ such that
\begin{equation}
p_{X, Y, Z} = r_{X,Y}q_{Z|Y}. \label{statement1CorollaryMC}
\end{equation}
 Then
\begin{equation}
(X\rightarrow Y\rightarrow Z)_{p_{X,Y,Z}} \label{statement3CorollaryMC}
\end{equation}
forms a Markov chain. In addition,
\begin{equation}
p_{Z|Y}=q_{Z|Y} \label{statement2CorollaryMC}.
\end{equation}
\end{Proposition}
\begin{IEEEproof}The proof of \eqref{statement3CorollaryMC} is contained \cite[Proposition~2.5]{Yeung08Book}. In addition, \eqref{statement2CorollaryMC} follows from \eqref{statement1CorollaryMC}.
\end{IEEEproof}
\subsection{Deterministic MMN Dominated by Product Distribution} \label{sectionDeterministicNetwork}
\subsubsection{Problem Formulation and Main Result}
\begin{Definition} \label{definitionDeterministicMatrix}
A conditional distribution $q_{Y|X}$ is said to be \textit{deterministic} if for each $x^*\in \mathcal{X}$, there exists a $y^*\in \mathcal{Y}$ such that $q_{Y|X}(y^*|x^*)=1$.
\end{Definition}
\smallskip
 \begin{Definition}\label{definitionDeterministicMMN}
 The MMN $(\mathcal{X}_\mathcal{I}, \mathcal{Y}_\mathcal{I}, \alpha, \boldsymbol{\mathcal{S}}, \boldsymbol{\mathcal{G}}, \boldsymbol{q})$ is said to be \textit{deterministic} if $q^{(h)}_{Y_{\mathcal{G}_h}|X_{\mathcal{S}^{h}}, Y_{\mathcal{S}^{h-1}}}$ is deterministic for each $h\in\{1, 2, \ldots, \alpha\}$.
 \end{Definition}
  \smallskip
 With the help of the following definition, we can completely characterize the capacity region for a class of deterministic MMNs with zero-delay nodes.
\smallskip
\begin{Definition} \label{defMMNdominatedByProduct}
The deterministic MMN $(\mathcal{X}_\mathcal{I}, \mathcal{Y}_\mathcal{I}, \alpha, \boldsymbol{\mathcal{S}}, \boldsymbol{\mathcal{G}}, \boldsymbol{q})$ is said to be \textit{dominated by product distributions} if the following holds for each distribution $p_{X_{\mathcal{I}}}$: \\
\indent Define $s_{X_i}$ to be the marginal distribution of $p_{X_{\mathcal{I}}}$ for each $i\in\mathcal{I}$, i.e., $s_{X_i}(x_i) = \sum_{x_j:j\in \mathcal{I} \setminus \{i\}}p_{X_{\mathcal{I}}}(x_{\mathcal{I}})$ for all $x_i$. In addition, define $
p_{X_{\mathcal{I}}, Y_{\mathcal{I}}}\triangleq p_{X_{\mathcal{I}}} \prod_{h=1}^\alpha q_{Y_{\mathcal{G}_h} | X_{\mathcal{S}^h}, Y_{\mathcal{G}^{h-1}}}^{(h)}
$
 and
$
 s_{X_{\mathcal{I}}, Y_{\mathcal{I}}}\triangleq (\prod_{i=1}^N s_{X_i})( \prod_{h=1}^\alpha q_{Y_{\mathcal{G}_h} | X_{\mathcal{S}^h}, Y_{\mathcal{G}^{h-1}}}^{(h)})$. Then for any $T\subseteq \mathcal{I}$, $H_{p_{X_{\mathcal{I}}, Y_{\mathcal{I}}}}(Y_{T^c}| X_{T^c}) \le H_{s_{X_{\mathcal{I}}, Y_{\mathcal{I}}}}(Y_{T^c}| X_{T^c})$.
\end{Definition}
\smallskip

The following is our main result in this section.
\smallskip
\begin{Theorem} \label{theoremDeterministicChannels}
Let $(\mathcal{X}_\mathcal{I}, \mathcal{Y}_\mathcal{I}, \alpha, \boldsymbol{\mathcal{S}}, \boldsymbol{\mathcal{G}}, \boldsymbol{q})$ be a deterministic MMN dominated by product distributions, and let
\begin{align}
\mathcal{R}_{\text{in}}^{\text{det}} & \triangleq   \bigcup_{\substack{p_{X_\mathcal{I}, Y_\mathcal{I}}: p_{X_\mathcal{I}, Y_\mathcal{I}}= \\ (\prod_{i=1}^N p_{X_i})(\prod_{h=1}^\alpha q_{Y_{\mathcal{G}_h}|X_\mathcal{S}^{h}, Y_\mathcal{G}^{h-1}}^{(h)})}} \bigcap_{T\subseteq \mathcal{I}: T^c \cap \mathcal{D} \ne \emptyset } \left\{ R_\mathcal{I}\left| \: \parbox[c]{2 in}{$ \sum_{ i\in T}  R_{i}
 \le  H_{p_{X_\mathcal{I},Y_\mathcal{I}}}(Y_{T^c}|X_{T^c}), \\
 R_i=0 \text{ for all }i\in\mathcal{V}^c$} \right.\right\}. \label{RinDet}
\end{align}
Then,
\begin{equation}
\mathcal{C}= \mathcal{C}_+ = \mathcal{R}_{\text{in}}^{\text{det}} \label{theoremDeterministicChannelsSt1}
\end{equation}
and hence the network is delay-independent. In particular, \eqref{theoremDeterministicChannelsSt1} holds for the deterministic relay network with no interference in \cite{multicastCapacityRelayNetworks} and the finite-field linear deterministic network in \cite{AvestimehrDeterministic,linearFiniteField09}, which implies that they are delay-independent.
\end{Theorem}
\smallskip

\begin{Remark}
It has been shown in \cite{multicastCapacityRelayNetworks} that the capacity region of the deterministic relay network with no interference is contained in the classical cut-set bound even though the network contains zero-delay nodes. Therefore, it is intuitive that the capacity region of any deterministic MMN with zero-delay nodes should be contained in the classical cut-set bound. In addition, the cut-set bound can be achieved if the deterministic MMN is dominated by product distributions. Combining the intuition and the fact provided above, it is intuitive that Theorem~\ref{theoremDeterministicChannels} should hold.
\end{Remark}
\medskip
\begin{Example}
\begin{figure}[!t]
 \centering
  \includegraphics[width=2.2 in, height=0.9 in, bb = 205 305 604 475, angle=0, clip=true]{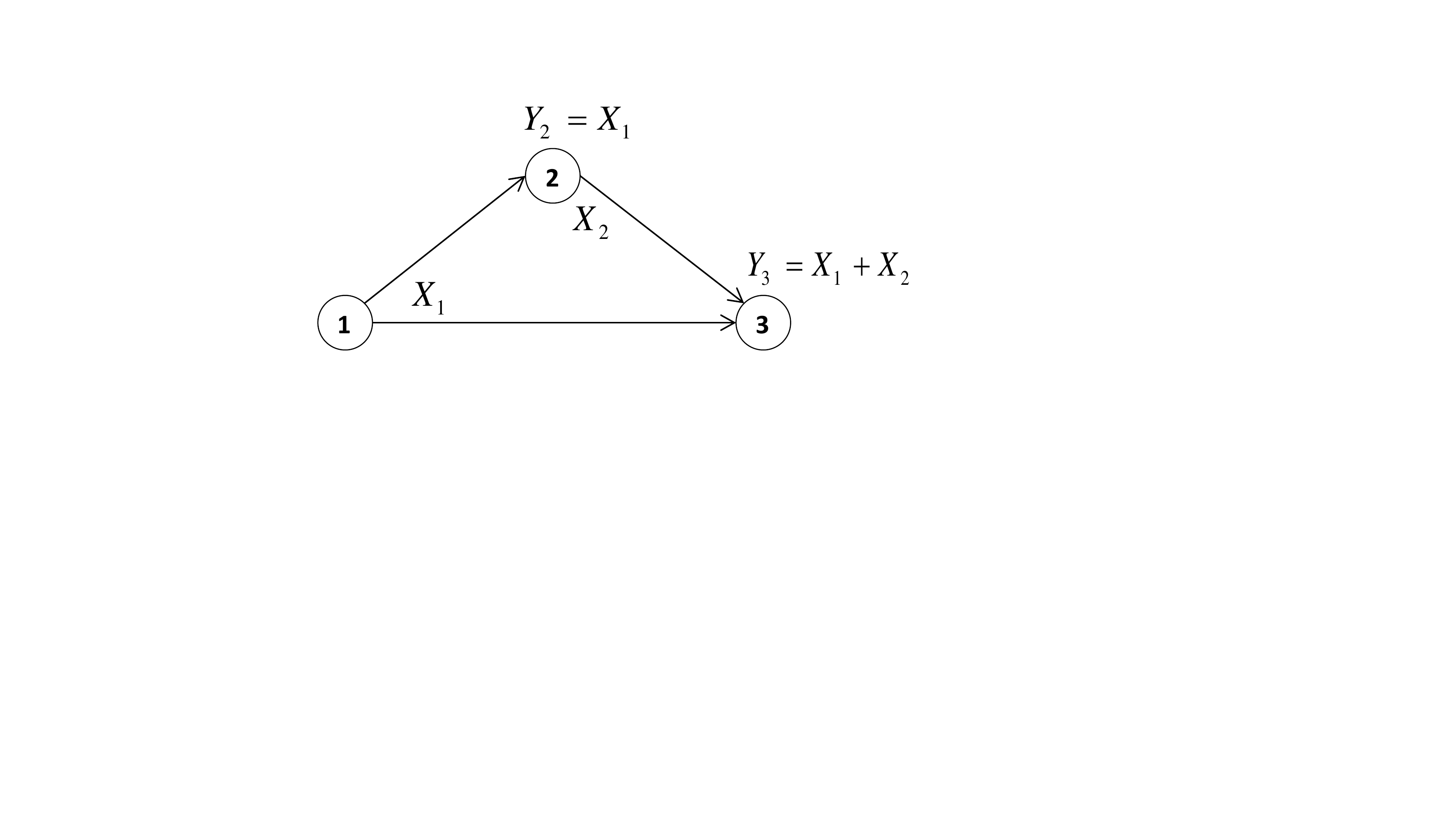}
\caption{A relay channel under the finite-field linear deterministic model.}  \label{RClinearDet}
\end{figure}
Consider a relay channel that consists of three nodes, where node~1 wants to transmit information to node~3 via a relay node~2. In each time slot, node~$i$ transmits $X_i$ and receives $Y_i$ for each~$i\in\{1,2,3\}$. All the alphabets are assumed to be binary, and we assume that $Y_2=X_1$ and $Y_3=X_1+X_2$. This relay channel is illustrated in Figure \ref{RClinearDet}. The relay channel is a finite-field linear deterministic network \cite{AvestimehrDeterministic}, and it can be formulated as a deterministic MMN with zero-delay nodes by setting $\boldsymbol{\mathcal{S}}\triangleq (\{1\},\{2,3\})$ and $\boldsymbol{\mathcal{G}}\triangleq (\{2\},\{1,3\})$ and choosing appropriate $q^{(1)}_{Y_2|X_1}$ and $q^{(2)}_{Y_1, Y_3|X_1, X_2, X_3, Y_2}$ such that $Y_2=X_1$ and $Y_3=X_1+X_2$ with probability one. Since node~2 incurs no delay under this formulation, we cannot characterize the capacity region by applying the classical cut-set bound. However, since every finite-field linear deterministic network is dominated by product distributions \cite[Section II-A]{noisyNetworkCoding}, Theorem~\ref{theoremDeterministicChannels} implies that this relay channel with a zero-delay node is delay-independent and its capacity region coincides with the classical cut-set bound. \hfill $\blacksquare$
\end{Example}
\medskip

 In the following, we provide the proof of Theorem~\ref{theoremDeterministicChannels}. Since the last statement of the theorem follows from the fact that the deterministic relay network with no interference and the finite-field linear deterministic network are dominated by product distributions \cite[Section II-A]{noisyNetworkCoding}, it suffices to prove \eqref{theoremDeterministicChannelsSt1}. To this end, it suffices to prove the achievability statement
 \begin{equation}
 \mathcal{R}_{\text{in}}^{\text{det}} \subseteq \mathcal{C}_+ \label{achievabilityStatementDet}
  \end{equation}
  and the converse statement
   \begin{equation}
 \mathcal{C}\subseteq \mathcal{R}_{\text{in}}^{\text{det}}. \label{converseStatementDet}
  \end{equation}
\subsubsection{Achievability}
In this subsection, we would like to show \eqref{achievabilityStatementDet} by using Theorem~\ref{theoremNNC} and Definition~\ref{definitionDeterministicMMN}.
 Since $\mathcal{R}_{\text{in}} \subseteq \mathcal{C}_+$ by Theorem~\ref{theoremNNC} (cf.\ \eqref{Rin}), it suffices to prove that
 \begin{equation}
 \mathcal{R}_{\text{in}}=\mathcal{R}_{\text{in}}^{\text{det}}. \label{theoremSt1lemmaDeterministicRin}
 \end{equation}
 Fix any $p_{X_\mathcal{I}, Y_\mathcal{I}}$ that satisfies
 \begin{equation}
 p_{X_\mathcal{I}, Y_\mathcal{I}}= \left(\prod_{i=1}^N p_{X_i}\right)\left(\prod_{h=1}^\alpha q_{Y_{\mathcal{G}_h}|X_\mathcal{S}^{h}, Y_\mathcal{G}^{h-1}}^{(h)}\right).  \label{theoremSt2lemmaDeterministicRin}
 \end{equation}
%
  Since
    \begin{equation*}
 p_{Y_\mathcal{I}|X_\mathcal{I}}= \prod_{h=1}^\alpha q_{Y_{\mathcal{G}_h}|X_\mathcal{S}^{h}, Y_\mathcal{G}^{h-1}}^{(h)} 
  \end{equation*}
  by \eqref{theoremSt2lemmaDeterministicRin} and $q_{Y_{\mathcal{G}_h} | X_{\mathcal{S}^h}, Y_{\mathcal{G}^{h-1}}}^{(h)}$ is deterministic for each $h\in\{1, 2, \ldots, \alpha\}$, it follows that $p_{Y_{\mathcal{I}}|X_{\mathcal{I}}}$ is deterministic and hence
 \[
 H_{p_{Y_{\mathcal{I}}|X_{\mathcal{I}}}}(Y_\mathcal{I}|X_\mathcal{I})=0,
 \]
 which then implies that
 \begin{equation}
 I_{p_{X_\mathcal{I},Y_\mathcal{I}}}(X_T; Y_{T^c}|X_{T^c})-H_{p_{X_\mathcal{I},Y_\mathcal{I}}}(Y_T|X_\mathcal{I}, Y_{T^c}) = H_{p_{X_\mathcal{I},Y_\mathcal{I}}}(Y_{T^c}|X_{T^c}). \label{lemmaCutsetDetinProofEq1}
 \end{equation}
Consequently, \eqref{theoremSt1lemmaDeterministicRin} follows from \eqref{Rin}, \eqref{RinDet} and \eqref{lemmaCutsetDetinProofEq1}.
\subsubsection{Converse}
In this subsection, we will show \eqref{converseStatementDet}. Given a $(B, n, M_{\mathcal{I}\times\mathcal{I}})$-code on the deterministic MMN and the messages $W_{\mathcal{I}}$, a careful inspection of Definitions~\ref{defCode},~\ref{defDiscreteMemoryless} and~\ref{definitionDeterministicMMN} will reveal that $(X_\mathcal{I}^n, Y_\mathcal{I}^n)$ is just a function of $W_{\mathcal{I}}$, which is formally stated in the following lemma. Since the proof of the lemma is straightforward, it is relegated to Appendix~\ref{appendixA}.
\smallskip
\begin{Lemma}\label{cutsetMCLemma*}
Let $(\mathcal{X}_\mathcal{I}, \alpha, \mathcal{Y}_\mathcal{I}, \boldsymbol{\mathcal{S}}, \boldsymbol{\mathcal{G}}, \boldsymbol{q})$ be a deterministic MMN. For any $(B, n, M_{\mathcal{I}\times\mathcal{I}})$-code on the network,
\begin{equation}
H_{p_{ W_{\mathcal{I}},X_{\mathcal{I}}^n, Y_{\mathcal{I}}^n  }}(X_{\mathcal{I}}^n, Y_{\mathcal{I}}^n |  W_{\mathcal{I}})=0, \label{lemmaStatement1MCLemma*}
\end{equation}
where $p_{W_{\mathcal{I}},X_{\mathcal{I}}^n, Y_{\mathcal{I}}^n  }$ is the joint distribution induced by the code according to Definitions~\ref{defCode} and~\ref{defDiscreteMemoryless}.
\end{Lemma}
\smallskip

In order to show that the capacity region of the deterministic MMN with zero-delay nodes lies within the classical cut-set bound, we will prove in Theorem~\ref{equivalentNetwork}, the theorem following the proposition below, that the deterministic MMN with zero-delay nodes is equivalent to some classical deterministic MMN. The following proposition is an important step for proving Theorem~\ref{equivalentNetwork}.
\smallskip
\begin{Proposition} \label{PropositionequivalentNetwork}
Let $(\mathcal{X}_\mathcal{I}, \mathcal{Y}_\mathcal{I}, \alpha, \boldsymbol{\mathcal{S}}, \boldsymbol{\mathcal{G}}, \boldsymbol{q})$ be a deterministic MMN. For any $(B, n, M_{\mathcal{I}\times\mathcal{I}})$-code on the network, if some $u$, $x_{\mathcal{I}}$ and $y_{\mathcal{I}}$ satisfy
\begin{equation}
\Pr\{U^{k-1} = u^{k-1}, X_{\mathcal{I}, k} = x_\mathcal{I}\}>0 \label{assumption0}
 \end{equation}
 and
 \begin{equation}
 \Pr\{U^{k-1} = u^{k-1}, X_{\mathcal{I}, k} = x_{\mathcal{I}},  Y_{\mathcal{I},k} = y_{\mathcal{I}} \}=0, \label{contradiction1}
  \end{equation}
then there exists some $h\in\{1, 2, \ldots, \alpha\}$ such that
$
q^{(h)}(y_{\mathcal{G}_h} | x_{\mathcal{S}^h}, y_{\mathcal{G}^{h-1}})=0
$
 (where $x_{\mathcal{S}^h}$ is a subtuple of $x_{\mathcal{I}}$, and $y_{\mathcal{G}_h}$ is a subtuple of $y_{\mathcal{I}}$).
\end{Proposition}
\begin{IEEEproof}
Suppose there exist $u$, $x_{\mathcal{I}}$ and $y_{\mathcal{I}}$ that satisfy \eqref{assumption0} and \eqref{contradiction1}. We prove the proposition by assuming the contrary. Assume
\begin{equation}
q^{(h)}(y_{\mathcal{G}_h} | x_{\mathcal{S}^h}, y_{\mathcal{G}^{h-1}} )>0 \label{assumption1}
\end{equation} for all $h\in\{1, 2, \ldots, \alpha\}$.
 We now prove by induction on $h$ that
\begin{equation}
\Pr\{U^{k-1}\! = u^{k-1}, X_{\mathcal{S}^{h},k} \! = x_{\mathcal{S}^{h}}, \! Y_{\mathcal{G}^{h},k}\!=y_{\mathcal{G}^{h}} \}>0 \label{assumption2}
\end{equation}
for each $h\in\{1, 2, \ldots, \alpha\}$. For $h=1$, the LHS of \eqref{assumption2} is
\begin{align}
& \Pr\{U^{k-1}\! = u^{k-1}, X_{\mathcal{S}^{1},k} \! = x_{\mathcal{S}^{1}}, \! Y_{\mathcal{G}^{1},k}\!=y_{\mathcal{G}^{1}} \} \notag \\
& \quad \stackrel{\text(a)}{=} p_{U^{k-1}, X_{\mathcal{S}^1,k}}(u^{k-1}, x_{\mathcal{S}^1}) q^{(1)}(y_{\mathcal{G}_1} | x_{\mathcal{S}^1}) \notag \\
& \quad \stackrel{\text{(b)}}{>} 0, \label{inductionFirstStatement}
\end{align}
where
\begin{enumerate}
\item[(a)] follows from Definitions~\ref{defDiscreteMemoryless} and~\ref{definitionDeterministicMMN}.
 \item[(b)] follows from \eqref{assumption0} and \eqref{assumption1}.
 \end{enumerate}
If \eqref{assumption2} holds for $h=m$, i.e.,
\begin{equation}
\Pr\{U^{k-1}\! = u^{k-1}, X_{\mathcal{S}^{m},k} \! = x_{\mathcal{S}^{m}}, \! Y_{\mathcal{G}^{m},k}\!=y_{\mathcal{G}^{m}} \}>0, \label{assumption2*}
\end{equation}
 then for $h=m+1$ such that $m+1 \le \alpha$,
\begin{align}
& \Pr\{U^{k-1} = u^{k-1}, X_{\mathcal{S}^{m+1},k} =x_{\mathcal{S}^{m+1}}, Y_{\mathcal{G}^{m+1},k}=y_{\mathcal{G}^{m+1}} \}\notag \\
&\stackrel{\text{(a)}}{=} p_{U^{k-1}, X_{\mathcal{S}^{m+1},k},Y_{\mathcal{G}^{m},k}}(u^{k-1}, x_{\mathcal{S}^{m+1}})   q^{(m+1)}(y_{\mathcal{G}_{m+1}} | x_{\mathcal{S}^{m+1}},y_{\mathcal{G}^{m}})\notag \\
&=p_{U^{k-1}, X_{\mathcal{S}^{m},k}, Y_{\mathcal{G}^{m},k}}(u^{k-1}, x_{\mathcal{S}^{m}}, y_{\mathcal{G}^{m}})  q^{(m+1)}(y_{\mathcal{G}_{m+1}} | x_{\mathcal{S}^{m+1}},y_{\mathcal{G}^{m}})
\notag \\
 & \quad \:\: p_{X_{\mathcal{S}_{m+1},k}|U^{k-1},X_{\mathcal{S}^{m},k}, Y_{\mathcal{G}^{m},k}}(x_{\mathcal{S}_{m+1}}|u^{k-1}, x_{\mathcal{S}^{m}},y_{\mathcal{G}^{m}})\notag \\
&\stackrel{\text{(b)}}{=} p_{U^{k-1}, X_{\mathcal{S}^{m},k}, Y_{\mathcal{G}^{m},k}}(u^{k-1}, x_{\mathcal{S}^{m}}, y_{\mathcal{G}^{m}})  q^{(m+1)}(y_{\mathcal{G}_{m+1}} | x_{\mathcal{S}^{m+1}},y_{\mathcal{G}^{m}})
\notag \\
 & \quad \:\: p_{X_{\mathcal{S}_{m+1},k}|U^{k-1}}(x_{\mathcal{S}_{m+1}}|u^{k-1}) \notag \\
 &\stackrel{\text{(c)}}{>} 0, \label{inductionStatement}
\end{align}
where
\begin{enumerate}
\item[(a)] follows from Definitions~\ref{defDiscreteMemoryless} and~\ref{definitionDeterministicMMN}.
\item[(b)] follows from Lemma~\ref{cutsetMCLemma*} that $(X_{\mathcal{S}^{m},k}, Y_{\mathcal{G}^{m},k})$ is a function of $U^{k-1}$.
    \item[(c)] follows from \eqref{assumption2*}, \eqref{assumption1} and \eqref{assumption0}.
\end{enumerate}
Consequently, it follows from \eqref{inductionFirstStatement}, \eqref{assumption2*} and \eqref{inductionStatement} that \eqref{assumption2} holds for $h=\alpha$ by mathematical induction, which then implies that
$
\Pr\{U^{k-1} = u^{k-1}, X_{\mathcal{S}^{\alpha},k}  = x_{\mathcal{S}^{\alpha}}, Y_{\mathcal{G}^{\alpha},k}=y_{\mathcal{G}^{\alpha}} \}>0$,
which contradicts \eqref{contradiction1}.
\end{IEEEproof}
\smallskip
Surprisingly, each deterministic MMN with zero-delay nodes is equivalent to some classical deterministic MMN, which is proved as follows.
\smallskip
\begin{Theorem} \label{equivalentNetwork}
For any $(B, n, M_{\mathcal{I}\times\mathcal{I}})$-code, the deterministic MMN specified by $(\mathcal{X}_\mathcal{I}, \mathcal{Y}_\mathcal{I}, \alpha, \boldsymbol{\mathcal{S}}, \boldsymbol{\mathcal{G}}, \boldsymbol{q})$ is equivalent to the deterministic MMN specified by $(\mathcal{X}_\mathcal{I}, \mathcal{Y}_\mathcal{I}, 1, \mathcal{I}, \mathcal{I}, q^{(1)} q^{(2)} \ldots q^{(\alpha)})$.
\end{Theorem}
\begin{IEEEproof}
Fix a $(B, n, M_{\mathcal{I}\times\mathcal{I}})$-code, and let $U^{k-1}\triangleq (W_{\mathcal{I}}, X_{\mathcal{I}}^{k-1}, Y_{\mathcal{I}}^{k-1})$ be the collection of random variables that are generated before the $k^{\text{th}}$ time slot. To prove the theorem statement for this code, it suffices to show that the following two statements are equivalent for each $k\in\{1, 2, \ldots, n\}$ (cf.\ \eqref{memorylessStatement}):\\
\textbf{Statement 1:} For each $h\in\{1, 2, \ldots, \alpha\}$,
\begin{align}
&  \Pr\{U^{k-1} = u^{k-1} , X_{\mathcal{S}^h, k} = x_{\mathcal{S}^h}, Y_{\mathcal{G}^h, k} = y_{\mathcal{G}^h}\} \notag\\
&\quad =
  \Pr\{U^{k-1} = u^{k-1} ,  X_{\mathcal{S}^h, k} = x_{\mathcal{S}^h}, Y_{\mathcal{G}^{h-1}, k} = y_{\mathcal{G}^{h-1}}\} q^{(h)}(y_{\mathcal{G}_h} | x_{\mathcal{S}^h}, y_{\mathcal{G}^{h-1}}). \label{memorylessStatement1}
\end{align}
\textbf{Statement 2:}
\begin{align}
&  \Pr\{U^{k-1} = u^{k-1} , X_{\mathcal{I}, k} = x_\mathcal{I}, Y_{\mathcal{I}, k} = y_\mathcal{I}\} \notag\\*
 & \:= \Pr\{U^{k-1}\! = u^{k-1} , X_{\mathcal{I}, k}\! = x_\mathcal{I}\}\prod_{h=1}^\alpha  q^{(h)}(y_{\mathcal{G}_h} | x_{\mathcal{S}^h}, y_{\mathcal{G}^{h-1}}). \label{memorylessStatement2}
\end{align}
Fix a $(B, n, M_{\mathcal{I}\times\mathcal{I}})$-code and a $k\in\{1, 2, \ldots, n\}$.
We first show that \eqref{memorylessStatement1} implies \eqref{memorylessStatement2}.
Suppose \eqref{memorylessStatement1} holds for each $h\in\{1, 2, \ldots, \alpha\}$. Consider the following three mutually exclusive cases:\\ \textbf{Case {$\Pr\{U^{k-1} = u^{k-1}, X_{\mathcal{I}, k} = x_\mathcal{I}\}=0$}:} \\
\indent
Both the LHS and the RHS of \eqref{memorylessStatement2} equal zero.  \\
 \textbf{Case {$\Pr\{U^{k-1} = u^{k-1}, X_{\mathcal{I}, k} = x_\mathcal{I}\}>0$} and\\ \text{\hspace{0.1 in}} {$\Pr\{U^{k-1} = u^{k-1}, X_{\mathcal{I}, k} = x_{\mathcal{I}},  Y_{\mathcal{I},k} = y_{\mathcal{I}} \}=0$}:}
\\
\indent For this case, the LHS of \eqref{memorylessStatement2} equals zero. By Proposition~\ref{PropositionequivalentNetwork}, there exists some $h\in\{1, 2, \ldots, \alpha\}$ such that $q^{(h)}(y_{\mathcal{G}_h} | x_{\mathcal{S}^h}, y_{\mathcal{G}^{h-1}})=0$, which implies that the RHS of \eqref{memorylessStatement2} equals zero.  \\
 \textbf{Case {$\Pr\{U^{k-1} = u^{k-1}, X_{\mathcal{I}, k} = x_{\mathcal{I}},  Y_{\mathcal{I}, k} = y_{\mathcal{I}}\}>0$}:}
 \\ \indent
For this case,
\begin{align*}
&  \Pr\{U^{k-1} = u^{k-1}, X_{\mathcal{I}, k} = x_\mathcal{I}, Y_{\mathcal{I}, k}  = y_\mathcal{I}\} \\
&\quad=  p_{U^{k-1}, X_{\mathcal{I},k}}(u^{k-1}, x_{\mathcal{I}})\prod_{h=1}^\alpha p_{Y_{\mathcal{G}_h,k} |U^{k-1} , X_{\mathcal{I},k}, Y_{\mathcal{G}^{h-1},k}}(y_{\mathcal{G}_h} |u^{k-1} , x_{\mathcal{I}}, y_{\mathcal{G}^{h-1}}) \\
&\quad \stackrel{\text{(a)}}{=} p_{U^{k-1}, X_{\mathcal{I},k}}(u^{k-1}, x_{\mathcal{I}}) \prod_{h=1}^\alpha p_{Y_{\mathcal{G}_h,k} |U^{k-1} , X_{\mathcal{S}^h,k}, Y_{\mathcal{G}^{h-1},k}}(y_{\mathcal{G}_h} |u^{k-1} , x_{\mathcal{S}^h}, y_{\mathcal{G}^{h-1}}) \\
&\quad  \stackrel{\eqref{memorylessStatement1}}{=} p_{U^{k-1}, X_{\mathcal{I},k}}(u^{k-1}, x_{\mathcal{I}}) \prod_{h=1}^\alpha  q^{(h)}(y_{\mathcal{G}_h} | x_{\mathcal{S}^h}, y_{\mathcal{G}^{h-1}}),
\end{align*}
where (a) follows from Lemma~\ref{cutsetMCLemma*} that for the $(B, n, M_{\mathcal{I}\times\mathcal{I}})$-code, $X_{\mathcal{I},k}$ is a function of $U^{k-1}$.
Therefore, the LHS and the RHS of \eqref{memorylessStatement2} are equal.

Combining the three mutually exclusive cases, we obtain that \eqref{memorylessStatement1} implies \eqref{memorylessStatement2}.
We now show that \eqref{memorylessStatement2} implies \eqref{memorylessStatement1}.
Suppose \eqref{memorylessStatement2} holds. Then for each $h\in\{1, 2, \ldots, \alpha\}$ and each $m\in\{1, 2, \ldots, h\}$,
\begin{align}
&  \Pr\{U^{k-1} = u^{k-1}, X_{\mathcal{S}^h,k} =x_{\mathcal{S}^h}, Y_{\mathcal{G}^{m},k}=y_{\mathcal{G}^{m}} \} \notag \\
&\quad = \sum\limits_{\substack{x_{\mathcal{S}_{h+1}},\ldots, x_{\mathcal{S}_{\alpha}}\\ y_{\mathcal{G}_{m+1}}, \ldots, y_{\mathcal{G}_{\alpha}}}}  p_{U^{k-1}, X_{\mathcal{I}, k}, Y_{\mathcal{I},k}}(u^{k-1}, x_{\mathcal{I}}, y_{\mathcal{I}}) \notag  \\
&\quad \stackrel{\eqref{memorylessStatement2}}{=} \sum\limits_{\substack{x_{\mathcal{S}_{h+1}},\ldots, x_{\mathcal{S}_{\alpha}}\\ y_{\mathcal{G}_{m+1}}, \ldots, y_{\mathcal{G}_{\alpha}}}}  p_{U^{k-1}, X_{\mathcal{I}, k}}(u^{k-1}, x_{\mathcal{I}})\prod_{\ell=1}^\alpha  q^{(\ell)}(y_{\mathcal{G}_\ell} | x_{\mathcal{S}^\ell}, y_{\mathcal{G}^{\ell-1}} )\notag   \\
&\quad = \sum_{x_{\mathcal{S}_{h+1}},\ldots, x_{\mathcal{S}_{\alpha}}}  p_{U^{k-1}, X_{\mathcal{I}, k}}(u^{k-1}, x_{\mathcal{I}}) \prod_{\ell=1}^{m}  q^{(\ell)}(y_{\mathcal{G}_\ell} | x_{\mathcal{S}^\ell}, y_{\mathcal{G}^{\ell-1}} ) \notag \\
&\quad \stackrel{\text{(a)}}{=} p_{U^{k-1}, X_{\mathcal{S}^h, k}}(u^{k-1}, x_{\mathcal{S}^h}) \prod_{\ell=1}^{m}  q^{(\ell)}(y_{\mathcal{G}_\ell} | x_{\mathcal{S}^\ell}, y_{\mathcal{G}^{\ell-1}} ), \label{equivalentNetworkEqn1}
\end{align}
where (a) follow from the fact that $m\le h$.
Then, for each $h\in\{1, 2, \ldots, \alpha\}$, the equality in \eqref{memorylessStatement1} can be verified by substituting \eqref{equivalentNetworkEqn1} into the LHS and the RHS.
\end{IEEEproof}
\smallskip
The following lemma simplifies the outer bound in Theorem~\ref{thmCapacityRegionMulticast} for the deterministic MMN.
\smallskip
 \begin{Lemma}\label{lemmaCutsetDeterministicChannelsRout}
 Let $(\mathcal{X}_\mathcal{I}, \mathcal{Y}_\mathcal{I}, \alpha, \boldsymbol{\mathcal{S}}, \boldsymbol{\mathcal{G}}, \boldsymbol{q})$ be a deterministic MMN. Define
 \begin{align}
\mathcal{R}_{\text{out}}^{\text{det}} & \triangleq   \bigcup_{\substack{p_{X_{\mathcal{I}}, Y_{\mathcal{I}}}:p_{X_{\mathcal{I}}, Y_{\mathcal{I}}}=\\  p_{X_\mathcal{I}}\prod_{h=1}^\alpha q_{Y_{\mathcal{G}_h} | X_{\mathcal{S}^h}, Y_{\mathcal{G}^{h-1}}}^{(h)}}} \bigcap_{T\subseteq \mathcal{I}: T^c \cap \mathcal{D} \ne \emptyset }
&  \left\{ R_\mathcal{I}\left| \: \parbox[c]{2 in}{$ \sum_{ i\in T}  R_{i}
 \le  H_{p_{X_{\mathcal{I}}, Y_{\mathcal{I}}}}(Y_{T^c}| X_{T^c}),\\
 R_i=0 \text{ for all }i\in\mathcal{V}^c$} \right.\right\}. \label{RoutDet}
\end{align}
  Then,
  \[
  \mathcal{C}\subseteq  \mathcal{R}_{\text{out}}^{\text{det}}.
  \]
 \end{Lemma}
 \begin{IEEEproof}
Suppose $R_{\mathcal{I}}$ is an achievable rate tuple for $(\mathcal{X}_\mathcal{I}, \mathcal{Y}_\mathcal{I}, \alpha, \boldsymbol{\mathcal{S}}, \boldsymbol{\mathcal{G}}, \boldsymbol{q})$. It follows from Definition~\ref{defAchievableRate} that there exists a sequence of $(B, n, M_{\mathcal{I}})$-codes on $(\mathcal{X}_\mathcal{I}, \mathcal{Y}_\mathcal{I}, \alpha, \boldsymbol{\mathcal{S}}, \boldsymbol{\mathcal{G}}, \boldsymbol{q})$ such that
$
 \lim\limits_{n\rightarrow \infty} \frac{\log M_{i}}{n} \ge R_{i}
$
for each $i\in\mathcal{I}$ and
$
 \lim\limits_{n\rightarrow \infty} P_{\text{err}}^{n} = 0$.
Since $(\mathcal{X}_\mathcal{I}, \mathcal{Y}_\mathcal{I}, \alpha, \boldsymbol{\mathcal{S}}, \boldsymbol{\mathcal{G}}, \boldsymbol{q})$ is equivalent to $(\mathcal{X}_\mathcal{I}, \mathcal{Y}_\mathcal{I}, 1, \mathcal{I}, \mathcal{I}, q^{(1)} q^{(2)} \ldots q^{(\alpha)})$ for any $(B, n, M_{\mathcal{I}\times\mathcal{I}})$-code on the deterministic DMN by Theorem~\ref{equivalentNetwork}, it follows from Definition~\ref{defCode} that $R_{\mathcal{I}}$ is achievable for \linebreak $(\mathcal{X}_\mathcal{I}, \mathcal{Y}_\mathcal{I}, 1, \mathcal{I}, \mathcal{I}, q^{(1)} q^{(2)} \ldots q^{(\alpha)})$, which then implies from Theorem~\ref{thmCapacityRegionMulticast} that there exists some $p_{X_{\mathcal{I}}, Y_{\mathcal{I}}}^*$ satisfying
\begin{equation}
p_{X_{\mathcal{I}}, Y_{\mathcal{I}}}^*=p_{X_\mathcal{I}}^*\prod_{h=1}^\alpha q_{Y_{\mathcal{G}_h} | X_{\mathcal{S}^h}, Y_{\mathcal{G}^{h-1}}}^{(h)} \label{lemmaStatement1CutsetDeterministic}
 \end{equation}
 such that for any $T\subseteq\mathcal{I}$ such that $T^c\cap \mathcal{D} \ne \emptyset$,
\begin{align}
\sum_{i\in T}  R_{i} & \le I_{p_{X_{\mathcal{I}}, Y_{\mathcal{I}}}^*}(X_T;Y_{T^c}| X_{T^c}) \notag\\
& =  H_{p_{X_{\mathcal{I}}, Y_{\mathcal{I}}}^*}(Y_{T^c}| X_{T^c})  - H_{p_{X_{\mathcal{I}}, Y_{\mathcal{I}}}^*}(Y_{T^c}| X_{\mathcal{I}}) \notag\\
& \stackrel{\text{(a)}}{=} H_{p_{X_{\mathcal{I}}, Y_{\mathcal{I}}}^*}(Y_{T^c}| X_{T^c}) \label{lemmaStatement2CutsetDeterministic}
\end{align}
where (a) follows from the fact that
 \[
 p_{ Y_{\mathcal{I}}|X_{\mathcal{I}}}^* \stackrel{\eqref{lemmaStatement1CutsetDeterministic}}{=} \prod_{h=1}^\alpha q_{Y_{\mathcal{G}_h} | X_{\mathcal{S}^h}, Y_{\mathcal{G}^{h-1}}}^{(h)}
 \]
 is deterministic. Consequently, it follows from \eqref{RoutDet}, \eqref{lemmaStatement1CutsetDeterministic} and \eqref{lemmaStatement2CutsetDeterministic} that $R_{\mathcal{I}}\in\mathcal{R}_{\text{out}}^{\text{det}}$.
 \end{IEEEproof}
\medskip

We are now ready to prove \eqref{converseStatementDet} as follows. Using Lemma~\ref{lemmaCutsetDeterministicChannelsRout}, we obtain
$\mathcal{C}\subseteq \mathcal{R}_{\text{out}}^{\text{det}}$
  where $\mathcal{R}_{\text{out}}^{\text{det}}$ is defined in \eqref{RoutDet}. In addition, it follows from \eqref{RinDet}, \eqref{RoutDet} and Definition~\ref{defMMNdominatedByProduct} that $\mathcal{R}_{\text{out}}^{\text{det}}\subseteq \mathcal{R}_{\text{in}}^{\text{det}}$. Consequently, $\mathcal{C}\subseteq \mathcal{R}_{\text{in}}^{\text{det}}$.

\subsection{MMN Consisting of Independent DMCs} \label{sectionDM-MMNconsistingOfDMCs}
\subsubsection{Problem Formulation and Main Result}
 Consider a DM-MMN $(\mathcal{X}_\mathcal{I}, \mathcal{Y}_\mathcal{I}, \alpha, \boldsymbol{\mathcal{S}}, \boldsymbol{\mathcal{G}}, \boldsymbol{q})$ defined as follows: The edge set of the network is characterized by
  \begin{equation}
 \Omega \triangleq \bigcup_{h=1}^\alpha \mathcal{S}^h\times \mathcal{G}_h, \label{defOmega}
 \end{equation}
and a DMC denoted by $q_{Y_{i,j}|X_{i,j}}$ is associated with every edge $(i,j)\in \Omega$, where $\mathcal{X}_{i,j}$ and $\mathcal{Y}_{i,j}$ are the input and output alphabets of the DMC carrying information from node~$i$ to node~$j$. The definition of $\Omega$ in \eqref{defOmega} ensures that
$q_{ Y_{\mathcal{G}_h}|X_{\mathcal{S}^h}, Y_{\mathcal{G}^{h-1}}}^{(h)}$
  can be well-defined for each $h\in\{1, 2,\ldots, \alpha\}$.
  For each $(i,j)\in \Omega$, the capacity of channel $q_{Y_{i,j}|X_{i,j}}$, denoted by $C_{i,j}$, is attained by some $\bar p_{X_{i,j}}$ due to the channel coding theorem, i.e.,
 \begin{align}
 C_{i,j}& \triangleq \max_{p_{X_{i,j}}}I_{p_{X_{i,j}} q_{Y_{i,j}|X_{i,j}}}(X_{i,j};Y_{i,j}) \
 \notag\\
 &= I_{ \bar p_{X_{i,j}} q_{Y_{i,j}|X_{i,j}}}(X_{i,j};Y_{i,j}). \label{defPTPCapacity}
 \end{align}
 For all the other $(\tilde i, \tilde j)\in \Omega^c$, we assume without loss of generality that
 \begin{equation}
 \mathcal{X}_{\tilde i,\tilde j}=\mathcal{Y}_{\tilde i,\tilde j}=\{0\} \label{alphabetSize=1}
 \end{equation}
 and $C_{\tilde i,\tilde j}=0$.
Then, we define the input and output alphabets for each node $i$ in the following natural way:
\begin{equation}
\mathcal{X}_i \triangleq \mathcal{X}_{i,1}\times \mathcal{X}_{i,2} \times \ldots \times \mathcal{X}_{i,N} \label{alphabetXSequence}
\end{equation}
and
\begin{equation}
\mathcal{Y}_i \triangleq \mathcal{Y}_{1,i}\times \mathcal{Y}_{2,i} \times \ldots \times \mathcal{Y}_{N,i}\label{alphabetYSequence}
\end{equation}
for each $i\in\mathcal{I}$. In addition, we define
\begin{equation}
q_{Y_{\mathcal{G}_h}|X_{\mathcal{S}^h},Y_{\mathcal{G}^{h-1}}}^{(h)} \triangleq \prod_{(i,j)\in\mathcal{S}^h \times \mathcal{G}_h}q_{Y_{i,j}|X_{i,j}} \label{defPTPnetworkChannel}
\end{equation}
for each $h\in\{1, 2, \ldots, \alpha\}$,
i.e., the random transformations (noises) from $X_{i,j}$ to $Y_{i,j}$  are independent and each channel $q_{Y_{\mathcal{G}_h}|X_{\mathcal{S}^h},Y_{\mathcal{G}^{h-1}}}^{(h)}$ is in a product form.
We call the network described above the \textit{DM-MMN consisting of independent DMCs}. The classical MMN consisting of independent DMCs studied in \cite{networkEquivalencePartI} is a special case of this network model when $\alpha=1$ and $\Omega=\mathcal{I}\times \mathcal{I}$. The following is the main result in this section, and the proof will be presented in the next two subsections.
\medskip
\begin{Theorem} \label{thmCapacityPTP}
For the DM-MMN consisting of independent DMCs, define
\begin{equation}
\mathcal{R}_{\text{in}}^{\text{DMCs}} \triangleq \bigcap_{T\subseteq \mathcal{I}: T^c \cap \mathcal{D} \ne \emptyset } \left\{ R_\mathcal{I}\left| \: \parbox[c]{1.8 in}{$ \sum_{ i\in T} R_{i}
 \le  \sum_{ (i,j)\in T \times T^c} C_{i,j},\\
 R_i=0 \text{ for all }i\in\mathcal{V}^c$} \right.\right\}. \label{RDMCs}
\end{equation}
Then,
\[
\mathcal{C} = \mathcal{C}_+ = \mathcal{R}_{\text{in}}^{\text{DMCs}}
\]
 and hence the network is delay-independent.
\end{Theorem}
\medskip

\begin{Remark}
In network coding theory, it is well-known that the classical cut-set bound (also called \textit{max-flow bound}) always holds for networks consisting of noiseless bit-pipes with zero-delay nodes \cite[Chapter 18]{Yeung08Book}. Therefore, it is intuitive that by replacing the noiseless bit-pipes by independent DMCs, the cut-set bound still serves as an outer bound on the capacity region. On the other hand, the cut-set bound can be achieved for MMNs consisting of independent DMCs. Combining the intuition and the fact provided above, it is intuitive that Theorem~\ref{thmCapacityPTP} should hold.
\end{Remark}
\medskip
\begin{Example} \label{exampleRelayIndDMCs}
\begin{figure}[!t]
 \centering
  \includegraphics[width=2 in, height=1 in, bb = 205 287 516 446, angle=0, clip=true]{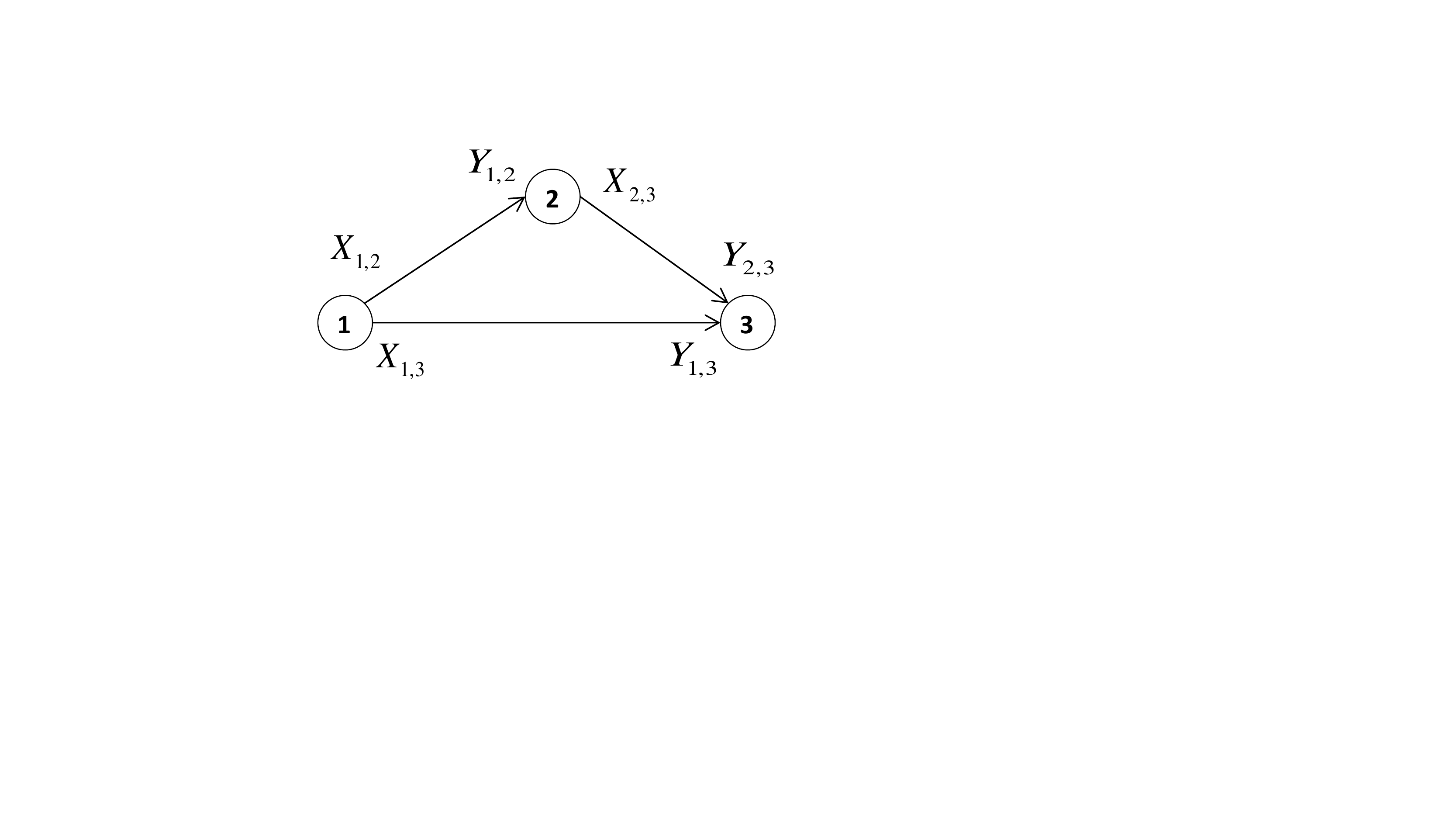}
\caption{A relay channel consisting of independent DMCs.}  \label{RCindDMCs}
\end{figure}
Consider a relay channel that consists of three nodes and three edges connecting the nodes, where node~1 wants to transmit information to node~3 via a relay node~2 through edges $(1,2)$, $(1,3)$ and $(2,3)$. In each time slot, node~$i$ transmits $X_{i,j}$ to node~$j$ through edge~$(i,j)$  and receives $Y_{\ell,i}$ from node~$\ell$ through edge $(\ell, i)$. Each edge is associated with a DMC. The three DMCs associated with the three edges, denoted by $q_{Y_{1,2}| X_{1,2}}$,  $q_{Y_{1,3}| X_{1,3}}$ and $q_{Y_{2,3}| X_{2,3}}$ respectively, are assumed to be independent, i.e.,
\begin{equation*}
p_{Y_{1,2}, Y_{1,3}, Y_{2,3}| X_{1,2}, X_{1,3}, X_{2,3}} = q_{Y_{1,2}| X_{1,2}}  q_{Y_{1,3}| X_{1,3}} q_{Y_{2,3}| X_{2,3}} 
\end{equation*}
regardless of the distribution of $(X_{1,2}, X_{1,3}, X_{2,3})$.
This relay channel is illustrated in Figure \ref{RCindDMCs}. The relay channel can be formulated as a DM-MMN consisting of independent DMCs by setting $\boldsymbol{\mathcal{S}}\triangleq (\{1\},\{2,3\})$, $\boldsymbol{\mathcal{G}}\triangleq (\{2\},\{1,3\})$, $\Omega \triangleq (\mathcal{S}_1\times \mathcal{G}_1) \cup (\mathcal{S}^2 \times \mathcal{G}_2)$, $X_1\triangleq (X_{1,2}, X_{1,3})$, $X_2\triangleq X_{2,3}$, $Y_2\triangleq Y_{1,2}$, $Y_3\triangleq (Y_{1,3}, Y_{2,3})$, $q_{Y_2|X_1}^{(1)}\triangleq q_{Y_{1,2}|X_{1,2}}$ and $q_{Y_3|X_1, X_2}^{(2)}\triangleq q_{Y_{1,3}|X_{1,3}}q_{Y_{2,3}|X_{2,3}}$. The set of non-trivial edges $\{(1,2), (1,3), (2,3)\}$ is inside $\Omega$ by \eqref{defOmega}. Since node~2 incurs no delay under this formulation, we cannot characterize the capacity region by applying the classical cut-set bound. Surprisingly, Theorem~\ref{thmCapacityPTP} implies that this relay channel with a zero-delay node is delay-independent and its capacity region coincides with the classical cut-set bound. \hfill $\blacksquare$
%
\end{Example}

 In the following, we provide the proof of Theorem~\ref{thmCapacityPTP}. To this end, it suffices to prove the achievability statement
 \begin{equation}
 \mathcal{R}_{\text{in}}^{\text{DMCs}} \subseteq \mathcal{C}_+ \label{achievabilityStatementPTP}
  \end{equation}
  and the converse statement
   \begin{equation}
 \mathcal{C}\subseteq \mathcal{R}_{\text{in}}^{\text{DMCs}}. \label{converseStatementPTP}
  \end{equation}

\subsubsection{Achievability}
In this subsection, we would like to prove \eqref{achievabilityStatementPTP}.
Since the DMCs $q_{Y_{i,j}|X_{i,j}}$ are all independent and each of the DMC can carry information at a rate arbitrarily close to the capacity, it is intuitive that $\mathcal{R}_{\text{in}}^{\text{DMCs}}$ lies in the positive-delay region of the DM-MMN consisting of independent DMCs, which is proved as follows.

Let $(\mathcal{X}_\mathcal{I}, \mathcal{Y}_\mathcal{I}, \alpha, \boldsymbol{\mathcal{S}}, \boldsymbol{\mathcal{G}}, \boldsymbol{q})$ be the DM-MMN consisting of independent DMCs whose positive-delay region is denoted by $\mathcal{C}_+$, and construct a counterpart of the channel $(\mathcal{X}_\mathcal{I}, \mathcal{Y}_\mathcal{I}, q_{Y_\mathcal{I}|X_\mathcal{I}})$ as follows: Let $(\bar {\mathcal{X}}_\mathcal{I}, \bar {\mathcal{Y}}_\mathcal{I}, \alpha, \bar{\boldsymbol{\mathcal{S}}}, \bar {\boldsymbol{\mathcal{G}}}, \bar  {\boldsymbol{q}})$ be a noiseless DM-MMN consisting of independent DMCs with multicast demand $(\mathcal{V}, \mathcal{D})$ such that for each $(i,j)\in \mathcal{I}\times \mathcal{I}$, the DMC carrying information from node~$i$ to node~$j$ is an error-free (noiseless) channel, denoted by $\bar q_{\bar X_{i,j} | \bar X_{i,j}}$, with capacity $C_{i,j}$ (cf.\ \eqref{defPTPCapacity}). To be more precise, $\bar q_{\bar X_{i,j} | \bar X_{i,j}}$ can carry $\lfloor n C_{i,j} \rfloor$ error-free bits for each $(i,j)\in \mathcal{I}\times \mathcal{I}$ for $n$ uses of $(\bar{\mathcal{X}}_\mathcal{I}, \bar{\mathcal{X}}_\mathcal{I}, \bar q_{\bar X_\mathcal{I}|\bar X_\mathcal{I}})$.  Let $\bar {\mathcal{C}}_+$ denote the positive-delay region of $(\bar {\mathcal{X}}_\mathcal{I}, \bar {\mathcal{Y}}_\mathcal{I}, \alpha, \bar{\boldsymbol{\mathcal{S}}}, \bar {\boldsymbol{\mathcal{G}}}, \bar  {\boldsymbol{q}})$. Since the original as well as the counterpart DM-MMNs consist of independent DMCs, it follows from the network equivalence theory \cite{networkEquivalencePartI} that $\mathcal{C}_+ = \bar {\mathcal{C}}_+$.
In addition, it has been shown in \cite[Section II-A]{noisyNetworkCoding} that $\bar {\mathcal{C}}_+ =\mathcal{R}_{\text{in}}^{\text{DMCs}}$. Consequently, $\mathcal{C}_+ = \bar {\mathcal{C}}_+=\mathcal{R}_{\text{in}}^{\text{DMCs}}$ and \eqref{achievabilityStatementPTP} holds.
\subsubsection{Converse of Theorem~\ref{thmCapacityPTP}}
 In this subsection, we would like to prove~\eqref{converseStatementPTP}. Define
  \begin{align}
& \mathcal{R}_{\text{out}}^{\text{DMCs}} \triangleq \notag\\*
&    \bigcup_{\substack{p_{ X_\mathcal{I},  Y_\mathcal{I}}: p_{ X_\mathcal{I},  Y_\mathcal{I}}= \\ \prod_{h=1}^\alpha (p_{X_{\mathcal{S}_h}|X_{\mathcal{S}^{h-1}}, Y_{\mathcal{G}^{h-1}}} q^{(h)}_{Y_{\mathcal{G}_h} | X_{\mathcal{S}^h},  Y_{\mathcal{G}^{h-1}}}) }} \!\!\!\! \bigcap_{T\subseteq \mathcal{I}: T^c \cap \mathcal{D} \ne \emptyset } \left\{ R_\mathcal{I}\left| \: \parbox[c]{3 in}{$\sum_{i\in T}R_i\le\sum_{h=1}^\alpha  I_{p_{ X_{\mathcal{I}},  Y_{\mathcal{I}}}}( X_{T\cap \mathcal{S}^h},  Y_{T\cap \mathcal{G}^{h-1}} ; \\ { }\hspace{1.2 in}  Y_{T^c\cap \mathcal{G}_h}|  X_{T^c \cap \mathcal{S}^h},   Y_{T^c\cap \mathcal{G}^{h-1}}),\\
 R_i=0 \text{ for all }i\in\mathcal{V}^c$} \right.\right\}. \label{RoutDMCs}
\end{align}
It follows form Theorem~\ref{thmCapacityRegionMulticast} that $\mathcal{C}\subseteq  \mathcal{R}_{\text{out}}^{\text{DMCs}}$. Therefore, it remains to show that
\begin{equation}
\mathcal{R}_{\text{out}}^{\text{DMCs}} \subseteq \mathcal{R}_{\text{in}}^{\text{DMCs}}. \label{RoutSubsetRinDMC}
\end{equation}
For any $p_{ X_\mathcal{I},  Y_\mathcal{I}} = \prod_{h=1}^\alpha (p_{X_{\mathcal{S}_h}|X_{\mathcal{S}^{h-1}}, Y_{\mathcal{G}^{h-1}}} q^{(h)}_{Y_{\mathcal{G}_h} | X_{\mathcal{S}^h},  Y_{\mathcal{G}^{h-1}}}) $, it follows from \eqref{defPTPnetworkChannel} that
\begin{equation}
p_{ X_\mathcal{I},  Y_\mathcal{I}} = \prod_{h=1}^\alpha \left(  p_{X_{\mathcal{S}_h}|X_{\mathcal{S}^{h-1}}, Y_{\mathcal{G}^{h-1}}} \prod_{(i,j)\in\mathcal{S}^h \times \mathcal{G}_h}q_{Y_{i,j}|X_{i,j}}\right). \label{distributionToBeMarginalizedDMCs}
\end{equation}
Marginalizing \eqref{distributionToBeMarginalizedDMCs}, we have
 \begin{equation}
p_{ X_{\mathcal{S}^m},  Y_{\mathcal{G}^m}} = \prod_{h=1}^m \left(  p_{X_{\mathcal{S}_h}|X_{\mathcal{S}^{h-1}}, Y_{\mathcal{G}^{h-1}}} \prod_{(i,j)\in\mathcal{S}^h \times \mathcal{G}_h}q_{Y_{i,j}|X_{i,j}}\right) \label{distributionToBeMarginalizedDMCs***}
\end{equation}
  for each $m\in\{1, 2, \ldots, \alpha\}$.
Relabeling symbols in \eqref{distributionToBeMarginalizedDMCs***} and using \eqref{alphabetXSequence} and \eqref{alphabetYSequence}, we have
\[
p_{X_{\mathcal{S}^h\times \mathcal{I}}, Y_{\mathcal{I}\times \mathcal{G}^h}}=\prod_{\ell=1}^h \left(  p_{X_{\mathcal{S}_\ell \times \mathcal{I}}|X_{\mathcal{S}^{\ell-1}\times \mathcal{I}}, Y_{\mathcal{I}\times \mathcal{G}^{\ell-1}}} \prod_{(i,j)\in\mathcal{S}^\ell \times \mathcal{G}_\ell}q_{Y_{i,j}|X_{i,j}}\right)
\]
 for each $h\in\{1, 2, \ldots, \alpha\}$, which implies from Proposition~\ref{propositionMCsimplification} that for each $h\in\{1, 2, \ldots, \alpha\}$ and each $(i,j)\in \mathcal{S}^h \times \mathcal{G}_h$ (cf.\ \eqref{defOmega}),
\begin{equation}
( \{(X_{k, \ell}, Y_{k, \ell}): (k, \ell)\in \mathcal{S}^h\times \mathcal{G}^h, (k, \ell)\ne (i,j)\} \rightarrow X_{i,j} \rightarrow Y_{i,j})_{p_{ X_\mathcal{I},  Y_\mathcal{I}}} \label{markovChainPTPNetwork}
\end{equation}
forms a Markov chain.
Following \eqref{RoutDMCs}, we consider the following chain of inequalities for a fixed $p_{ X_\mathcal{I},  Y_\mathcal{I}} =  \prod_{h=1}^\alpha (p_{X_{\mathcal{S}_h}|X_{\mathcal{S}^{h-1}}, Y_{\mathcal{G}^{h-1}}}  q_{ Y_{\mathcal{G}_h}| X_\mathcal{S}^{h},  Y_\mathcal{G}^{h-1}}^{(h)})$ and a fixed $T\subseteq \mathcal{I}:$
\begin{align}
& \sum_{h=1}^\alpha  I_{p_{ X_{\mathcal{I}},  Y_{\mathcal{I}}}}( X_{T\cap \mathcal{S}^h},  Y_{T\cap \mathcal{G}^{h-1}} ;   Y_{T^c\cap \mathcal{G}_h}|  X_{T^c \cap \mathcal{S}^h},   Y_{T^c\cap \mathcal{G}^{h-1}}) \notag\\
& =\sum_{h=1}^\alpha \sum_{j\in T^c\cap \mathcal{G}_h} I_{p_{ X_{\mathcal{I}},  Y_{\mathcal{I}}}}( X_{T\cap \mathcal{S}^h},  Y_{T\cap \mathcal{G}^{h-1}} ;   Y_j|  X_{T^c \cap \mathcal{S}^h},   Y_{T^c\cap \mathcal{G}^{h-1}}, \{ Y_\ell\}_{\ell \in T^c\cap \mathcal{G}_h, \ell<j}) \notag\\
& \stackrel{\eqref{alphabetYSequence}}{=}  \sum_{h=1}^\alpha \hspace{-0.2 in}\sum_{\qquad i\in \mathcal{I}, j\in T^c\cap \mathcal{G}_h} \hspace{-0.2 in}I_{p_{ X_{\mathcal{I}},  Y_{\mathcal{I}}}}( X_{T\cap \mathcal{S}^h},  Y_{T\cap \mathcal{G}^{h-1}} ;   Y_{i,j}|  X_{T^c \cap \mathcal{S}^h},   Y_{T^c\cap \mathcal{G}^{h-1}}, \{ Y_\ell\}_{\ell \in T^c\cap \mathcal{G}_h, \ell<j}, \{ Y_{m,j}\}_{m \in \mathcal{I}, m<i}) \notag\\
& \stackrel{\text{(a)}}{=} \sum_{h=1}^\alpha \hspace{-0.2 in} \sum_{\qquad i\in \mathcal{S}^h, j\in T^c\cap \mathcal{G}_h} \hspace{-0.2 in}I_{p_{ X_{\mathcal{I}},  Y_{\mathcal{I}}}}( X_{T\cap \mathcal{S}^h},  Y_{T\cap \mathcal{G}^{h-1}} ;   Y_{i,j}|  X_{T^c \cap \mathcal{S}^h},   Y_{T^c\cap \mathcal{G}^{h-1}}, \{ Y_\ell\}_{\ell \in T^c\cap \mathcal{G}_h, \ell<j}, \{ Y_{m,j}\}_{m \in \mathcal{I}, m<i}) \notag\\
& \stackrel{\text{(b)}}{=} \sum_{h=1}^\alpha \hspace{-0.2 in} \sum_{\qquad i\in T\cap\mathcal{S}^h, j\in T^c\cap \mathcal{G}_h} \hspace{-0.2 in} I_{p_{ X_{\mathcal{I}},  Y_{\mathcal{I}}}}( X_{T\cap \mathcal{S}^h},  Y_{T\cap \mathcal{G}^{h-1}} ;   Y_{i,j}|  X_{T^c \cap \mathcal{S}^h},   Y_{T^c\cap \mathcal{G}^{h-1}}, \{ Y_\ell\}_{\ell \in T^c\cap \mathcal{G}_h, \ell<j}, \{ Y_{m,j}\}_{m \in \mathcal{I}, m<i}) \notag\\
& \stackrel{\text{(c)}}{\le} \sum_{h=1}^\alpha \hspace{-0.2 in} \sum_{\qquad i\in T\cap\mathcal{S}^h, j\in T^c\cap \mathcal{G}_h} \hspace{-0.2 in} I_{p_{ X_{\mathcal{I}},  Y_{\mathcal{I}}}}( X_{i,j} ;   Y_{i,j}) \notag\\
&\stackrel{\eqref{defPTPCapacity}}{\le} \sum_{h=1}^\alpha \hspace{-0.2 in} \sum_{\qquad i\in T\cap\mathcal{S}^h, j\in T^c\cap \mathcal{G}_h} \hspace{-0.2 in} C_{i,j} \notag\\
& \le \sum_{h=1}^\alpha \hspace{-0.2 in}\sum_{\qquad i\in T, j\in T^c\cap \mathcal{G}_h} \hspace{-0.2 in} C_{i,j} \notag\\
& = \sum_{(i,j)\in T\times T^c} C_{i,j}, \label{st1corollaryOuterBoundPTP}
\end{align}
where
\begin{enumerate}
\item[(a)] follows from the fact that for each $h\in\{1, 2, \ldots, \alpha\}$ and each $(i,j)\in (\mathcal{I}\setminus\mathcal{S}^h) \times \mathcal{G}_h$, $(i,j)$ lies in $\Omega^c$ (cf.~\eqref{defOmega}) and hence
\begin{align*}
& I_{p_{ X_{\mathcal{I}},  Y_{\mathcal{I}}}}( X_{T\cap \mathcal{S}^h},  Y_{T\cap \mathcal{G}^{h-1}} ;   Y_{i,j}|  X_{T^c \cap \mathcal{S}^h},   Y_{T^c\cap \mathcal{G}^{h-1}}, \{ Y_\ell\}_{\ell \in T^c\cap \mathcal{G}_h, \ell<j}, \{ Y_{m,j}\}_{m \in \mathcal{I}, m<i}) \notag\\
&\le H_{p_{ X_{\mathcal{I}},  Y_{\mathcal{I}}}}(Y_{i,j}) \notag\\
& \stackrel{\eqref{alphabetSize=1}}{=} 0.
\end{align*}
\item[(b)] follows from the fact that for each $h\in\{1, 2, \ldots, \alpha\}$ and each $(i,j)\in (T^c\cap \mathcal{S}^h) \times (T^c\cap \mathcal{G}_h)\subseteq \Omega$,
    \begin{align*}
& I_{p_{ X_{\mathcal{I}},  Y_{\mathcal{I}}}}( X_{T\cap \mathcal{S}^h},  Y_{T\cap \mathcal{G}^{h-1}} ;   Y_{i,j}|  X_{T^c \cap \mathcal{S}^h},   Y_{T^c\cap \mathcal{G}^{h-1}}, \{ Y_\ell\}_{\ell \in T^c\cap \mathcal{G}_h, \ell<j}, \{ Y_{m,j}\}_{m \in \mathcal{I}, m<i}) \notag\\
& =  H_{p_{ X_{\mathcal{I}},  Y_{\mathcal{I}}}}(Y_{i,j}|  X_{T^c \cap \mathcal{S}^h},   Y_{T^c\cap \mathcal{G}^{h-1}}, \{ Y_\ell\}_{\ell \in T^c\cap \mathcal{G}_h, \ell<j}, \{ Y_{m,j}\}_{m \in \mathcal{I}, m<i}) \notag\\
&\qquad - H_{p_{ X_{\mathcal{I}},  Y_{\mathcal{I}}}}( Y_{i,j}|  X_{\mathcal{S}^h},   Y_{T^c\cap \mathcal{G}^{h-1}}, \{ Y_\ell\}_{\ell \in T^c\cap \mathcal{G}_h, \ell<j}, \{ Y_{m,j}\}_{m \in \mathcal{I}, m<i},  Y_{T\cap \mathcal{G}^{h-1}}) \notag\\
&\stackrel{\eqref{markovChainPTPNetwork}}{=} H_{p_{ X_{\mathcal{I}},  Y_{\mathcal{I}}}}(Y_{i,j}|X_{i,j}) -  H_{p_{ X_{\mathcal{I}},  Y_{\mathcal{I}}}}(Y_{i,j}|X_{i,j})\notag\\
& = 0.
\end{align*}
\item[(c)] follows from the fact that for each $h\in\{1, 2, \ldots, \alpha\}$ and each $(i,j)\in (T\cap \mathcal{S}^h) \times (T^c\cap \mathcal{G}_h)\subseteq \Omega$,
    \begin{align*}
& I_{p_{ X_{\mathcal{I}},  Y_{\mathcal{I}}}}( X_{T\cap \mathcal{S}^h},  Y_{T\cap \mathcal{G}^{h-1}} ;   Y_{i,j}|  X_{T^c \cap \mathcal{S}^h},   Y_{T^c\cap \mathcal{G}^{h-1}}, \{ Y_\ell\}_{\ell \in T^c\cap \mathcal{G}_h, \ell<j}, \{ Y_{m,j}\}_{m \in \mathcal{I}, m<i}) \notag\\
& \le  H_{p_{ X_{\mathcal{I}},  Y_{\mathcal{I}}}}(Y_{i,j}) - H_{p_{ X_{\mathcal{I}},  Y_{\mathcal{I}}}}( Y_{i,j}|  X_{\mathcal{S}^h},   Y_{T^c\cap \mathcal{G}^{h-1}}, \{ Y_\ell\}_{\ell \in T^c\cap \mathcal{G}_h, \ell<j}, \{ Y_{m,j}\}_{m \in \mathcal{I}, m<i},  Y_{T\cap \mathcal{G}^{h-1}}) \notag\\
&\stackrel{\eqref{markovChainPTPNetwork}}{=} H_{p_{ X_{\mathcal{I}},  Y_{\mathcal{I}}}}(Y_{i,j}) - H_{p_{ X_{\mathcal{I}},  Y_{\mathcal{I}}}}(Y_{i,j}|X_{i,j})\notag\\
& = I_{p_{ X_{\mathcal{I}},  Y_{\mathcal{I}}}}(X_{i,j};Y_{i,j}).
\end{align*}
\end{enumerate}
Consequently, it follows from \eqref{RDMCs}, \eqref{RoutDMCs} and \eqref{st1corollaryOuterBoundPTP} that \eqref{RoutSubsetRinDMC} holds.

\subsection{Wireless Erasure Network} \label{sectionErasureNetworks}
\subsubsection{Problem Formulation and Main Result}
Consider a DM-MMN $(\mathcal{X}_\mathcal{I}, \mathcal{Y}_\mathcal{I}, \alpha, \boldsymbol{\mathcal{S}}, \boldsymbol{\mathcal{G}}, \boldsymbol{q})$ defined as follows: Similar to the MMN consisting of independent DMCs discussed in the previous section, we let
\begin{equation}
\Omega \triangleq \bigcup_{h=1}^\alpha \mathcal{S}^h \times \mathcal{G}_h
  \label{defOmegaWEN}
 \end{equation}
 characterize the edge set of the network so that
$q_{ Y_{\mathcal{G}_h}|X_{\mathcal{S}^h}, Y_{\mathcal{G}^{h-1}}}^{(h)}$ can be well-defined for each $h\in\{1, 2,\ldots, \alpha\}$. To simulate the broadcast nature of wireless networks, we assume that in every time slot, each node~$i$ broadcasts a symbol $X_i$ for each $i\in\mathcal{I}$ and we let $\mathcal{X}_i$ denote the finite alphabet of $X_{i}$. For each $(i,j)\in \Omega$, we assume that node~$j$ receives $X_{i}$ with erasure probability $\varepsilon_{i,j}\in [0,1]$, and we let $Y_{i,j}$ and
$
\mathcal{Y}_{i,j}\triangleq\mathcal{X}_i \cup \{\varepsilon\}
$
 denote the received symbol and its alphabet respectively where $\varepsilon$ denotes the erasure symbol.
  For every edge $(i^\prime, j^\prime)$ that is not in $\Omega$, we set its erasure probability $\varepsilon_{i^\prime, j^\prime}$ to be $1$ and
  \begin{equation}
  \mathcal{Y}_{i^\prime, j^\prime} \triangleq \{\varepsilon\}, \label{alphabetSize=1WEN}
   \end{equation}
   indicating that no information can be transmitted from $i^\prime$ to $j^\prime$. We let $q_{Y_{i,j}|X_{i}}$ characterize the channel corresponding to edge $(i,j)$ such that for each $x_i\in \mathcal{X}_i$ and each $y_{i,j}\in \mathcal{Y}_{i,j}$,
  \begin{equation}
   q_{Y_{i,j}|X_{i}}(y_{i,j}|x_{i}) = \begin{cases} 1-\varepsilon_{i,j} &\text{if $y_{i,j}=x_{i}$,}\\ \varepsilon_{i,j}& \text{if $y_{i,j}=\varepsilon$.}\end{cases} \label{PrYijGivenXij}
   \end{equation}
   The symbols transmitted on the edges in $\Omega$ are assumed to be erased independently, i.e.,
   \begin{equation}
   \Pr\left\{\left.\bigcap_{(i,j)\in \Omega}\{Y_{i,j}=y_{i,j}\}\right|X_{\mathcal{I}}=x_{\mathcal{I}}\right\} = \prod_{(i,j)\in \Omega}q_{Y_{i,j}|X_{i}}(y_{i,j}|x_{i}) \label{indErasureOmega}
   \end{equation}
   for each $x_{\mathcal{I}}\in \mathcal{X}_\mathcal{I}$ and each $|\Omega|$-dimensional tuple $(y_{i,j}: (i,j)\in \Omega)\in \prod_{(i,j)\in \Omega}\mathcal{Y}_{i,j}$.

   For each $(i,j)\in \mathcal{I}\times \mathcal{I}$, let $E_{i,j}$ and $\mathcal{E}_{i,j}$ be the indicator random variable for the erasure occurred on edge $(i,j)$ and its alphabet respectively such that
  \begin{equation}
  E_{i,j} \triangleq \mathbf{1}\left(\left\{ Y_{i,j}= \varepsilon \right\}\right) = \begin{cases}1 & \text{if $Y_{i,j}=\varepsilon$,} \\0 & \text{if $Y_{i,j}\ne \varepsilon$.} \end{cases}
  \label{defEij}
  \end{equation}
  Let
  \[
  E_{\mathcal{I}\times\mathcal{I}}\triangleq (E_{1,1}, E_{1,2}, \ldots, E_{N,N})
  \]
be the $N^2$-dimensional random tuple containing all the $E_{i,j}$'s so that $E_{\mathcal{I}\times\mathcal{I}}$ characterizes the \textit{network erasure pattern}, and let $\mathcal{E}_{\mathcal{I}\times\mathcal{I}}$ denote the alphabet of $E_{\mathcal{I}\times\mathcal{I}}$.
 Recalling that $(\mathcal{V}, \mathcal{D})$ is the multicast demand, we assume that the following two statements hold: \\
 \noindent (i) All the destinations are contained in $\mathcal{G}_\alpha$, i.e., $\mathcal{D}\subseteq \mathcal{G}_\alpha$. \\
  \noindent (ii) The network erasure pattern $E_{\mathcal{I}\times\mathcal{I}}$ in each time slot is accessible by each destination node in $\mathcal{D}$.
   \\
Since $E_{\mathcal{I}\times\mathcal{I}}$ is a function of $Y_{\mathcal{I}}$ by \eqref{defEij}, there exists some conditional distribution $\chi_{E_{\mathcal{I}\times\mathcal{I}}|Y_{\mathcal{I}}}$ such that
  \begin{equation}
\chi_{E_{\mathcal{I}\times \mathcal{I}}|Y_{\mathcal{I}}}(e_{\mathcal{I}\times \mathcal{I}}|y_{\mathcal{I}})  = \begin{cases} 1 & \text{if $e_{i,j}=\mathbf{1}(\{y_{i,j}=\varepsilon\})$ for all $(i,j)\in \mathcal{I}\times \mathcal{I}$,} \\
0& \text{otherwise}
\end{cases} \label{defChi}
  \end{equation}
   for all $y_{\mathcal{I}}$ and $e_{\mathcal{I}\times \mathcal{I}}$.
   We are now ready to formally define $\mathcal{X}_\mathcal{I}$, $\mathcal{Y}_\mathcal{I} $ and $\boldsymbol{q}$ as follows: For each $i\in\mathcal{I}$, recalling that $\mathcal{X}_i$ is the finite alphabet of the symbol $X_{i,k}$ broadcast by node~$i$, we define
 \begin{equation}
 \mathcal{X}_\mathcal{I} \triangleq \mathcal{X}_1\times \mathcal{X}_2 \times \ldots \times \mathcal{X}_N. \label{defWENalphabetX}
\end{equation}
Recalling that
$\mathcal{Y}_{i,j}= \mathcal{X}_i \cup \{\varepsilon\}$
is the alphabet of the noisy version of $X_{i,k}$ that is received by node~$j$ in each time slot for each $(i,j)\in \Omega$,
 we define for each $h\in \{1, 2, \ldots, \alpha\}$ and each $m\in\mathcal{G}_h$
 \begin{equation}
 \mathcal{Y}_m \triangleq
 \begin{cases}
 \left(\prod_{i\in \mathcal{I}}  \mathcal{Y}_{i,m}\right)\times  \mathcal{E}_{\mathcal{I}\times \mathcal{I}}& \text{if $m$ is an element in $\mathcal{D} \subseteq\mathcal{G}_\alpha$,} \\ \prod_{i\in \mathcal{S}^h}  \mathcal{Y}_{i,m} & \text{otherwise.}\end{cases}
 \label{defWENalphabetYm}
 \end{equation}
 The definition of $\mathcal{Y}_m$ in \eqref{defWENalphabetYm} is divided into two cases because we assume according to Statements~(i) and~(ii) that the destination nodes in $\mathcal{D}\subseteq\mathcal{G}_\alpha$ have access to the network erasure pattern.
  After defining $\mathcal{Y}_m$ for each $m\in \mathcal{I}$ in \eqref{defWENalphabetYm}, we define
\begin{equation}
 \mathcal{Y}_\mathcal{I} \triangleq \mathcal{Y}_1\times \mathcal{Y}_2 \times \ldots \times \mathcal{Y}_N. \label{defWENalphabetY}
\end{equation}
Based on the definitions of $\mathcal{X}_\mathcal{I}$ and $\mathcal{Y}_\mathcal{I}$ in \eqref{defWENalphabetX} and \eqref{defWENalphabetY} respectively and recalling \eqref{PrYijGivenXij} and \eqref{defChi}, we define $q_{Y_{\mathcal{G}_h}|X_{\mathcal{S}^h}, Y_{\mathcal{S}^{h-1}}}^{(h)}$ for each $h\in \{1,2, \ldots, \alpha\}$ as
\begin{align}
& q_{Y_{\mathcal{G}_h}|X_{\mathcal{S}^h}, Y_{\mathcal{S}^{h-1}}}^{(h)}(y_{\mathcal{G}_h}|x_{\mathcal{S}^h},y_{\mathcal{S}^{h-1}})\notag\\
&\triangleq \begin{cases}
\left(\prod_{i\in \mathcal{I}} \prod_{m\in \mathcal{G}_h} q_{Y_{i,m}|X_{i}}(y_{i,m}|x_{i})\right)\chi_{E_{\mathcal{I} \times \mathcal{I}}|Y_{\mathcal{I}}}(e_{\mathcal{I} \times \mathcal{I}}|y_{\mathcal{I}}) & \text{if $h=\alpha$,} \\ \prod_{i\in \mathcal{S}^h} \prod_{m\in \mathcal{G}_h} q_{Y_{i,m}|X_{i}}(y_{i,m}|x_{i}) & \text{otherwise} \end{cases}
\label{defYmGivenX}
 \end{align}
 for all $x_{\mathcal{S}^h}\in \mathcal{X}_{\mathcal{S}^h}$, $y_{\mathcal{G}^h}\in\mathcal{Y}_{\mathcal{G}^h}$ and $e_{\mathcal{I} \times \mathcal{I}}\in \mathcal{E}_{\mathcal{I} \times \mathcal{I}}$, where for each $m\in\mathcal{G}_h$
 \[
 y_m = \begin{cases} ((y_{i,m}: i\in\mathcal{I}),e_{\mathcal{I} \times \mathcal{I}}) &  \text{if $m$ is an element in $\mathcal{D} \subseteq\mathcal{G}_\alpha$,}\\ (y_{i,m}: i\in\mathcal{S}^h) & \text{otherwise.}\end{cases}
 \]
We call the network described above the \textit{wireless erasure network}. The random variables $X_{\mathcal{I}}$ and $Y_{\mathcal{I}}$ in the wireless erasure network are generated according to this order
  \[
 X_{\mathcal{S}_1}, Y_{\mathcal{G}_1}, X_{\mathcal{S}_2}, Y_{\mathcal{G}_2}, \ldots, X_{\mathcal{S}_\alpha}, Y_{\mathcal{G}_\alpha}
 \]
 (cf.\ \eqref{orderExplanation}), which implies from \eqref{defWENalphabetYm}, \eqref{defWENalphabetY} and \eqref{defYmGivenX} that $X_\mathcal{I}$, $\{Y_{i,j}\}_{(i,j)\in \Omega}$ and $E_{\mathcal{I}\times \mathcal{I}}$ are generated according to this order
  \begin{align}
  X_{\mathcal{S}_1}, \{Y_{i,j}\}_{(i,j)\in \mathcal{S}^1\times \mathcal{G}_1}, X_{\mathcal{S}_2}, \{Y_{i,j}\}_{(i,j)\in \mathcal{S}^2\times \mathcal{G}_2}, \ldots,X_{\mathcal{S}_\alpha}, \{Y_{i,j}\}_{(i,j)\in \mathcal{S}^\alpha\times \mathcal{G}_\alpha}, E_{\mathcal{I}\times \mathcal{I}}\,. \label{randomVariableSequenceWEN}
  \end{align}
 It may not be obvious from \eqref{randomVariableSequenceWEN} that $X_\mathcal{I}$ and $E_{\mathcal{I}\times \mathcal{I}}$ are always independent, but it follows from \eqref{PrYijGivenXij}, \eqref{indErasureOmega} and \eqref{defChi} that for any $e_{\mathcal{I}\times \mathcal{I}}\in \mathcal{E}_{\mathcal{I}\times \mathcal{I}}$ and $x_\mathcal{I}\in \mathcal{X}_\mathcal{I}$,
 \begin{equation}
 \Pr\{E_{\mathcal{I}\times \mathcal{I}}=e_{\mathcal{I}\times \mathcal{I}}|X_\mathcal{I}=x_\mathcal{I}\} = \prod_{(i,j)\in \Omega} \left(\varepsilon_{i,j}^{\mathbf{1}\{e_{i,j}=1\}}(1-\varepsilon_{i,j})^{\mathbf{1}\{e_{i,j}=0\}}\right), \label{EIIindependetOfXI*} 
 \end{equation}
 which implies the independence between $X_\mathcal{I}$ and $E_{\mathcal{I}\times \mathcal{I}}$, i.e.,
 \begin{equation}
 \Pr\{E_{\mathcal{I}\times \mathcal{I}}=e_{\mathcal{I}\times \mathcal{I}}|X_\mathcal{I}=x_\mathcal{I}\} = \Pr\{E_{\mathcal{I}\times \mathcal{I}}=e_{\mathcal{I}\times \mathcal{I}}\} \label{EIIindependetOfXI} 
 \end{equation}
for any $e_{\mathcal{I}\times \mathcal{I}}\in \mathcal{E}_{\mathcal{I}\times \mathcal{I}}$ and $x_\mathcal{I}\in \mathcal{X}_\mathcal{I}$. The classical wireless erasure network studied in \cite{dana06} is a special case of our model when $\alpha=1$ and $\Omega = \mathcal{I}\times \mathcal{I}$.
The following theorem is the main result in this section, and the proof will be provided in the next two subsections.
\medskip
\begin{Theorem}\label{thmWEN}
For the wireless erasure network, let
\begin{align}
\mathcal{R}_{\text{in}}^{\text{WEN}}& \triangleq  \bigcap_{T\subseteq \mathcal{I}: T^c \cap \mathcal{D} \ne \emptyset } \left\{ R_\mathcal{I}\left| \: \parbox[c]{2.3 in}{$ \sum\limits_{i\in  T}  R_i
 \le \sum_{i\in T}  \left(1-\prod_{j\in T^c}e_{i,j} \right)|\mathcal{X}_i|, \vspace{0.04 in}\\
 R_i=0 \text{ for all }i\in\mathcal{V}^c$} \right.\right\}. \label{defRWEN}
\end{align}
Then,
\[
\mathcal{C} = \mathcal{C}_+ =\mathcal{R}_{\text{in}}^{\text{WEN}}
\]
 and hence the network is delay-independent.
\end{Theorem}
\medskip

\begin{Remark}
For the wireless erasure network with zero-delay nodes, due to the independence nature among the erasures, the network can be intuitively viewed as a MMN consisting of independent erasure channels, whose capacity region is contained in the classical cut-set bound by Theorem~\ref{thmCapacityPTP}. On the other hand, it has been shown in \cite{dana06} that the cut-set bound can be achieved for the wireless erasure network. Combining the intuition and the fact provided above, it is intuitive that Theorem~\ref{thmWEN} should hold.
\end{Remark}
\medskip
\begin{Example} \label{exampleRelayWEN}
\begin{figure}[!t]
 \centering
  \includegraphics[width=2.4 in, height=0.9 in, bb = 204 304 646 456, angle=0, clip=true]{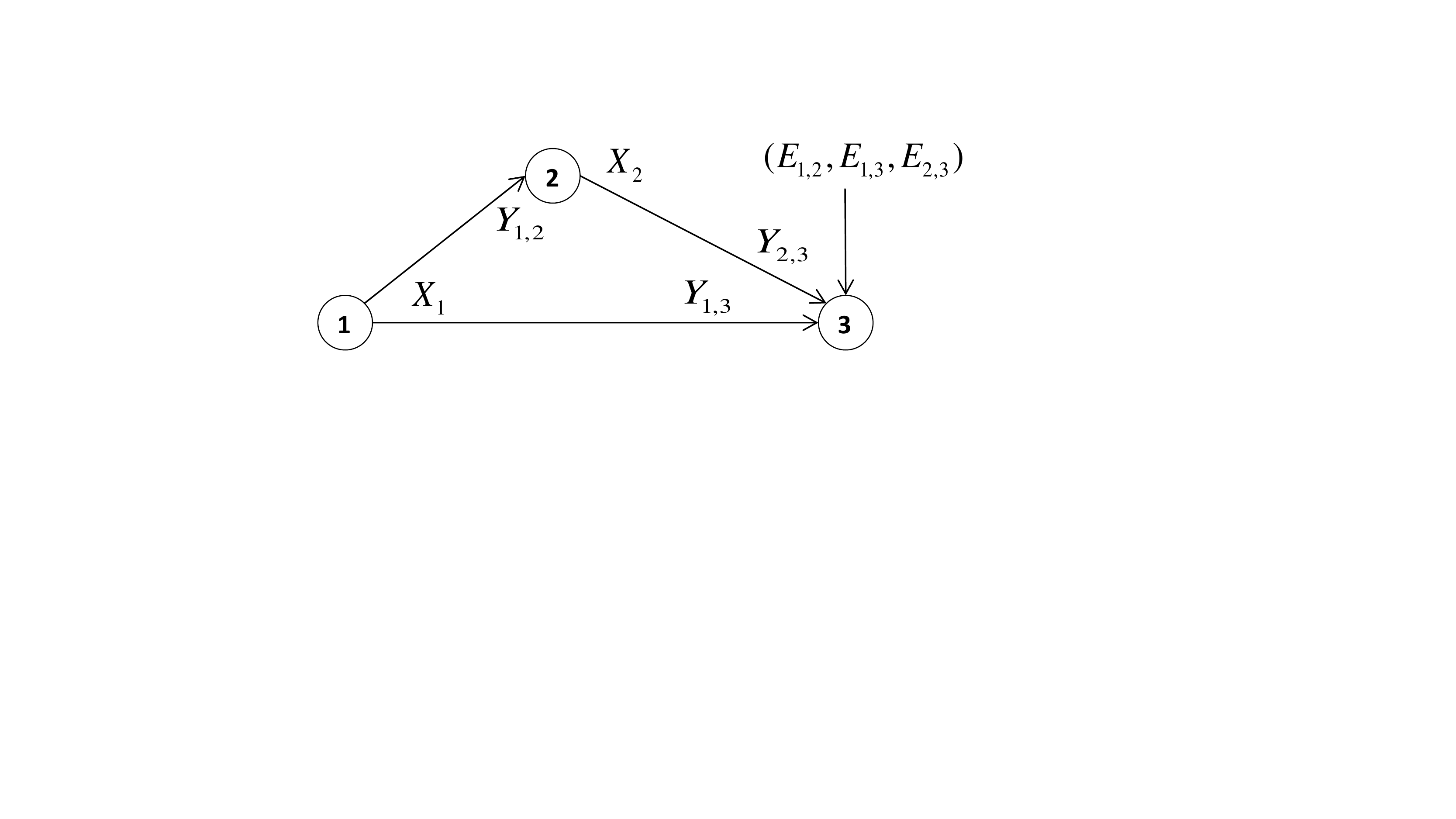}
\caption{A three-node wireless erasure network.}  \label{RCWEN}
\end{figure}
Consider a relay channel that consists of three nodes where node~1 wants to transmit information to node~3 via a relay node~2. In each time slot, node~$i$ transmits $X_i$ for each $i\in\{1,2,3\}$, while node~2 receives an erased version of $X_1$ denoted by $Y_{1,2}$ and node~3 receives erased versions of $X_1$ and~$X_2$ denoted by $Y_{1,3}$ and $Y_{2,3}$ respectively. Let $E_{i,j}$ denote the erasure random variable for $(i,j)$ where
\begin{equation}
E_{i,j} = \begin{cases} 0 & \text{if $X_i$ is not erased at node~$j$, i.e., $Y_{i,j}=X_i$,} \\ 1 & \text{otherwise, i.e., $X_i$ is erased at node~$j$.} \end{cases} \label{defEijInExample}
\end{equation}
The erasures are assumed to be independent, i.e.,
$p_{E_{1,2}, E_{1,3}, E_{2,3}}=p_{E_{1,2}} p_{E_{1,3}} p_{E_{2,3}}$
regardless of the distribution of $(X_1, X_2)$. In addition, node~3 is assumed to have access of the network erasure pattern $(E_{1,2}, E_{1,3}, E_{2,3})$ (note that $E_{1,3}$ and $E_{2,3}$ can be deduced from $Y_{1,3}$ and $Y_{2,3}$ respectively by \eqref{defEijInExample}, but $E_{1,2}$ is an extra information provided for node~3 for decoding).
This relay channel is illustrated in Figure~\ref{RCWEN}. The relay channel can be formulated as a wireless erasure network by setting $\boldsymbol{\mathcal{S}}\triangleq (\{1\},\{2,3\})$, $\boldsymbol{\mathcal{G}}\triangleq (\{2\},\{1,3\})$, $\Omega \triangleq (\mathcal{S}_1\times \mathcal{G}_1) \cup (\mathcal{S}^2 \times \mathcal{G}_2)$, $Y_2\triangleq Y_{1,2}$ and $Y_3\triangleq (Y_{1,3}, Y_{2,3}, E_{1,2}, E_{1,3}, E_{2,3})$. The set of non-trivial edges $\{(1,2), (1,3), (2,3)\}$ is contained in $\Omega$ by \eqref{defOmegaWEN}. Since node~2 incurs no delay under this formulation, we cannot characterize the capacity region by applying the classical cut-set bound. Surprisingly, Theorem~\ref{thmWEN} implies that this three-node wireless erasure network with a zero-delay node is delay-independent. \hfill $\blacksquare$
%
\end{Example}

 In the following, we provide the proof of Theorem~\ref{thmWEN}. To this end, it suffices to prove the achievability statement
 \begin{equation}
 \mathcal{R}_{\text{in}}^{\text{WEN}} \subseteq \mathcal{C}_+ \label{achievabilityStatementWEN}
  \end{equation}
  and the converse statement
   \begin{equation}
 \mathcal{C}\subseteq \mathcal{R}_{\text{in}}^{\text{WEN}}. \label{converseStatementWEN}
  \end{equation}

\subsubsection{Achievability} \label{sectionAchThmWEN}
In this subsection, we would like to prove \eqref{achievabilityStatementWEN}. Since the achievability statement \eqref{achievabilityStatementWEN} has been shown in \cite{dana06} under the classical model which considers no zero-delay nodes, \eqref{achievabilityStatementWEN} holds naturally under our generalized-delay model. For completeness, the proof of \eqref{achievabilityStatementWEN} under our generalized-delay model is provided in Appendix~\ref{appendixB}.

\subsubsection{Converse}\label{sectionCovThmWEN}
In this subsection, we would like to prove \eqref{converseStatementWEN}. We will first prove the following counterpart of Theorem~\ref{thmCapacityRegionMulticast} to show an outer bound on $\mathcal{C}$, and then show that the outer bound is contained in $\mathcal{R}_{\text{in}}^{\text{WEN}}$.
\medskip
\begin{Lemma} \label{thmCapacityRegionMulticastWEN}
 Let $(\mathcal{X}_\mathcal{I}, \mathcal{Y}_\mathcal{I}, \alpha, \boldsymbol{\mathcal{S}}, \boldsymbol{\mathcal{G}}, \boldsymbol q)$ be a wireless erasure network, and let
  \begin{align}
& \mathcal{R}_{\text{out}}^{\text{WEN}} \triangleq \notag\\*
&    \bigcup_{\substack{p_{ X_\mathcal{I},  Y_\mathcal{I}}: p_{ X_\mathcal{I},  Y_\mathcal{I}}= \\ \prod_{h=1}^\alpha (p_{X_{\mathcal{S}_h}|X_{\mathcal{S}^{h-1}}, Y_{\mathcal{G}^{h-1}}} q^{(h)}_{Y_{\mathcal{G}_h} | X_{\mathcal{S}^h},  Y_{\mathcal{G}^{h-1}}}) }}\bigcap_{T\subseteq \mathcal{I}: T^c \cap \mathcal{D} \ne \emptyset } \left\{ R_\mathcal{I}\left| \: \parbox[c]{3 in}{$\sum_{i\in T}R_i\le\sum_{h=1}^\alpha  I_{p_{ X_{\mathcal{I}},  Y_{\mathcal{I}}}}( X_{T\cap \mathcal{S}^h},  Y_{T\cap \mathcal{G}^{h-1}} ; \\ { }\hspace{1 in}  Y_{T^c\cap \mathcal{G}_h}|  X_{T^c \cap \mathcal{S}^h},   Y_{T^c\cap \mathcal{G}^{h-1}}, E_{\mathcal{I}\times \mathcal{I}}),\\
 R_i=0 \text{ for all }i\in\mathcal{V}^c$} \right.\right\} \label{RoutWEN}
\end{align}
where $E_{\mathcal{I}\times \mathcal{I}}$, the network erasure pattern, is a function of $Y_{\mathcal{I}}$ defined by \eqref{defEij}.
Then,
\[
\mathcal{C}\subseteq \mathcal{R}_{\text{out}}^{\text{WEN}}.
\]
\end{Lemma}
\begin{IEEEproof}
Let $R_{\mathcal{I}}$ be an achievable rate tuple for the wireless erasure network denoted by $(\mathcal{X}_\mathcal{I}, \mathcal{Y}_\mathcal{I}, \alpha, \boldsymbol{\mathcal{S}}, \boldsymbol{\mathcal{G}}, \boldsymbol q)$. Then, there exists a sequence of $(B, n, M_{\mathcal{I}})$-codes on the network such that
\begin{equation}
 \lim_{n\rightarrow \infty} \frac{\log M_{i}}{n} \ge R_{i} \label{thmTempEq1WEN}
\end{equation}
and
\begin{equation}
 \lim_{n\rightarrow \infty} P_{\text{err}}^n = 0 \label{thmTempEq2WEN}
\end{equation}
for each $i\in \mathcal{I}$.
Fix any $T\subseteq \mathcal{I}$ such that $T^c\cap \mathcal{D}\ne \emptyset$, and let $d$ denote a node in $T^c\cap \mathcal{D}$.
Fix a $(B, n, M_{\mathcal{I}})$-code and let $E_{\mathcal{I}\times \mathcal{I},k}$ denote the network erasure pattern occurred in time slot~$k$ for each $k\in\{1, 2, \ldots, n\}$. Then, we consider the following chain of inequalities:
\begin{align}
\sum_{i\in T} \log M_{i}
& \stackrel{\text{(a)}}{=} H(W_{T}|W_{T^c})\notag \\*
& \stackrel{\text{(b)}}{=} H(W_{T}|W_{T^c}, E_{\mathcal{I}\times \mathcal{I}}^n)\notag \\
&= I(W_{T}; Y_{T^c}^n|W_{T^c},E_{\mathcal{I}\times \mathcal{I}}^n) +
H(W_{T}|Y_{T^c}^n,W_{T^c},E_{\mathcal{I}\times \mathcal{I}}^n) \notag\\
&\le I(W_{T}; Y_{T^c}^n|W_{T^c},E_{\mathcal{I}\times \mathcal{I}}^n) + H(W_{T}|Y_d^n, W_{d},E_{\mathcal{I}\times \mathcal{I}}^n)\notag \\
 &\stackrel{\text{(c)}}{\le}   I(W_{T};  Y_{T^c}^n|W_{T^c},E_{\mathcal{I}\times \mathcal{I}}^n) + 1+ P_{\text{err}}^n \sum_{i\in T} \log M_{i},
\label{cutseteqnSet1WEN}
\end{align}
where
\begin{enumerate}
\item[(a)] follows from the fact that the $N$ messages $W_{1}, W_{2}, \ldots, W_{N}$ are independent.
\item[(b)] follows from the fact that $W_{\mathcal{I}}$ and $E_{\mathcal{I}\times \mathcal{I}}^n$ are independent.
\item[(c)] follows from Fano's inequality.
\end{enumerate}
Following similar procedures for proving Theorem 1 in \cite{fongYeung15}, we can show by using \eqref{thmTempEq1WEN}, \eqref{thmTempEq2WEN} and \eqref{cutseteqnSet1WEN} that there exists a joint distribution $p_{X_\mathcal{I}, Y_\mathcal{I}}$ which depends on the sequence of $(B, n, M_{\mathcal{I}})$-codes but not on $T$ such that
\[
  p_{X_{\mathcal{I}}, Y_{\mathcal{I}}} =
\prod_{h=1}^\alpha (p_{X_{\mathcal{S}_h}|X_{\mathcal{S}^{h-1}}, Y_{\mathcal{G}^{h-1}}}
q^{(h)}_{Y_{\mathcal{G}_h} | X_{\mathcal{S}^h},  Y_{\mathcal{G}^{h-1}}})
\]
and
\begin{equation}
\sum_{i\in T} R_i\le  \sum_{h=1}^\alpha  I_{p_{X_{\mathcal{I}}, Y_{\mathcal{I}}}}(X_{T\cap \mathcal{S}^h}, Y_{T\cap \mathcal{G}^{h-1}} ;  Y_{T^c\cap \mathcal{G}_h}| X_{T^c \cap \mathcal{S}^h},  Y_{T^c\cap \mathcal{G}^{h-1}}, E_{\mathcal{I}\times \mathcal{I}}). \label{stInTheoremCon}
\end{equation}
Since $p_{X_\mathcal{I}, Y_\mathcal{I}}$ depends on only the sequence of $(B, n, M_{\mathcal{I}})$-codes but not on $T$, \eqref{stInTheoremCon} holds for all $T\subseteq \mathcal{I}$ such that $T^c\cap \mathcal{D}\ne \emptyset$. This completes the proof.
\end{IEEEproof}
\medskip

Since
\begin{equation}
\mathcal{C}\subseteq  \mathcal{R}_{\text{out}}^{\text{WEN}} \label{converseWENByTheorem2}
 \end{equation}
 by Lemma~\ref{thmCapacityRegionMulticastWEN} and our goal is to prove $\mathcal{C}\subseteq \mathcal{R}_{\text{in}}^{\text{WEN}} $, it remains to show that
\begin{equation}
\mathcal{R}_{\text{out}}^{\text{WEN}} \subseteq \mathcal{R}_{\text{in}}^{\text{WEN}}. \label{RoutSubsetRinWEN}
\end{equation}
For any $p_{ X_\mathcal{I},  Y_\mathcal{I}} = \prod_{h=1}^\alpha (p_{X_{\mathcal{S}_h}|X_{\mathcal{S}^{h-1}}, Y_{\mathcal{G}^{h-1}}} q^{(h)}_{Y_{\mathcal{G}_h} | X_{\mathcal{S}^h},  Y_{\mathcal{G}^{h-1}}})$, it follows from \eqref{PrYijGivenXij} and \eqref{defEij} that for each $h\in\{1, 2, \ldots, \alpha\}$ and each $(i,j)\in \mathcal{S}^h \times \mathcal{G}_h$, $Y_{i,j}$ is a function of $(X_{i}, E_{\mathcal{I}\times \mathcal{I}})$ and hence
\begin{equation}
\left(\{(X_{k}, Y_{k, \ell}): (k, \ell)\in (\mathcal{S}^h \setminus \{i\})\times \mathcal{G}^h\} \rightarrow (X_{i}, E_{\mathcal{I}\times \mathcal{I}}) \rightarrow Y_{i,j}\right)_{p_{ X_\mathcal{I},  Y_\mathcal{I}}} \label{markovChainWENNetwork}
\end{equation}
forms a Markov chain. Following \eqref{RoutSubsetRinWEN} and \eqref{RoutWEN}, we fix $p_{ X_\mathcal{I},  Y_\mathcal{I}} =  \prod_{h=1}^\alpha (p_{X_{\mathcal{S}_h}|X_{\mathcal{S}^{h-1}}, Y_{\mathcal{G}^{h-1}}}  q_{ Y_{\mathcal{G}_h}| X_\mathcal{S}^{h},  Y_\mathcal{G}^{h-1}}^{(h)})$ and $T\subseteq \mathcal{I}$ such that $T^c \cap \mathcal{D} \ne \emptyset$, and we consider
\begin{align}
& \sum_{h=1}^\alpha  I_{p_{ X_{\mathcal{I}},  Y_{\mathcal{I}}}}( X_{T\cap \mathcal{S}^h},  Y_{T\cap \mathcal{G}^{h-1}} ;   Y_{T^c\cap \mathcal{G}_h}|  X_{T^c \cap \mathcal{S}^h},   Y_{T^c\cap \mathcal{G}^{h-1}}, E_{\mathcal{I}\times \mathcal{I}}) \notag\\*
& \stackrel{\eqref{defWENalphabetYm}}{=}\sum_{h=1}^\alpha \sum_{i\in \mathcal{S}^h} I_{p_{ X_{\mathcal{I}},  Y_{\mathcal{I}}}}( X_{T\cap \mathcal{S}^h},  Y_{T\cap \mathcal{G}^{h-1}} ;  Y_{\{i\}\times (T^c\cap \mathcal{G}_h)}|  X_{T^c \cap \mathcal{S}^h},   Y_{T^c\cap \mathcal{G}^{h-1}}, \{ Y_{\{\ell\}\times (T^c\cap \mathcal{G}_h})\}_{\ell<i},E_{\mathcal{I}\times \mathcal{I}}) \notag\\
& \stackrel{\text{(a)}}{=}  \sum_{h=1}^\alpha \sum_{i\in T\cap \mathcal{S}^h} I_{p_{ X_{\mathcal{I}},  Y_{\mathcal{I}}}}( X_{T\cap \mathcal{S}^h},  Y_{T\cap \mathcal{G}^{h-1}} ;  Y_{\{i\}\times (T^c\cap \mathcal{G}_h)}|  X_{T^c \cap \mathcal{S}^h},   Y_{T^c\cap \mathcal{G}^{h-1}}, \{ Y_{\{\ell\}\times (T^c\cap \mathcal{G}_h)}\}_{\ell<i}, E_{\mathcal{I}\times \mathcal{I}}) \notag\\
& \stackrel{\text{(b)}}{\le} \sum_{h=1}^\alpha \sum_{i\in T\cap\mathcal{S}^h} I_{p_{ X_{\mathcal{I}},  Y_{\mathcal{I}}}}( X_{i} ;    Y_{\{i\}\times (T^c\cap \mathcal{G}_h)}|  Y_{\{i\}\times (T^c\cap \mathcal{G}^{h-1})}, E_{\mathcal{I}\times \mathcal{I}}) \notag\\
& \stackrel{\text{(c)}}{=}  \sum_{h=1}^\alpha \sum_{i\in T\cap\mathcal{S}^h} H_{p_{ X_{\mathcal{I}},  Y_{\mathcal{I}}}}(    Y_{\{i\}\times (T^c\cap \mathcal{G}_h)}| Y_{\{i\}\times (T^c\cap \mathcal{G}^{h-1})}, E_{\mathcal{I}\times \mathcal{I}})\notag\\
& = \sum_{i\in T\cap\mathcal{S}^h} H_{p_{ X_{\mathcal{I}},  Y_{\mathcal{I}}}}(Y_{\{i\}\times T^c}|E_{\mathcal{I}\times \mathcal{I}})\notag\\
& \le \sum_{i\in T\cap\mathcal{S}^h} H_{p_{ X_{\mathcal{I}},  Y_{\mathcal{I}}}}(Y_{\{i\}\times T^c}|E_{\{i\}\times T^c})
 \label{eqn4WENCon}
\end{align}
where
\begin{enumerate}
\item[(a)] follows from the fact that for each $h\in\{1, 2, \ldots, \alpha\}$ and each $i\in T^c\cap \mathcal{S}^h$,
    \begin{align*}
& I_{p_{ X_{\mathcal{I}},  Y_{\mathcal{I}}}}( X_{T\cap \mathcal{S}^h},  Y_{T\cap \mathcal{G}^{h-1}} ;  Y_{\{i\}\times (T^c\cap \mathcal{G}_h)}|  X_{T^c \cap \mathcal{S}^h},   Y_{T^c\cap \mathcal{G}^{h-1}}, \{ Y_{\{\ell\}\times T^c\cap \mathcal{G}_h}\}_{\ell<i},E_{\mathcal{I}\times \mathcal{I}}) \notag\\
& =  H_{p_{ X_{\mathcal{I}},  Y_{\mathcal{I}}}}(Y_{\{i\}\times (T^c\cap \mathcal{G}_h)}|  X_{T^c \cap \mathcal{S}^h},   Y_{T^c\cap \mathcal{G}^{h-1}}, \{ Y_{\{\ell\}\times T^c\cap \mathcal{G}_h}\}_{\ell<i},E_{\mathcal{I}\times \mathcal{I}}) \notag\\
&\qquad - H_{p_{ X_{\mathcal{I}},  Y_{\mathcal{I}}}}(Y_{\{i\}\times (T^c\cap \mathcal{G}_h)}|  X_{\mathcal{S}^h},   Y_{\mathcal{G}^{h-1}}, \{ Y_{\{\ell\}\times T^c\cap \mathcal{G}_h}\}_{\ell<i},E_{\mathcal{I}\times \mathcal{I}}) \notag\\
&\stackrel{\eqref{markovChainWENNetwork}}{=} H_{p_{ X_{\mathcal{I}},  Y_{\mathcal{I}}}}(Y_{\{i\}\times (T^c\cap \mathcal{G}_h)}|X_i,E_{\mathcal{I}\times \mathcal{I}}) -  H_{p_{ X_{\mathcal{I}},  Y_{\mathcal{I}}}}(Y_{\{i\}\times (T^c\cap \mathcal{G}_h)}|X_i,E_{\mathcal{I}\times \mathcal{I}})\notag\\
& = 0.
\end{align*}
\item[(b)] follows from the fact that for each $h\in\{1, 2, \ldots, \alpha\}$ and each $i\in T\cap \mathcal{S}^h$,
    \begin{align*}
& I_{p_{ X_{\mathcal{I}},  Y_{\mathcal{I}}}}( X_{T\cap \mathcal{S}^h},  Y_{T\cap \mathcal{G}^{h-1}} ;  Y_{\{i\}\times (T^c\cap \mathcal{G}_h)}|  X_{T^c \cap \mathcal{S}^h},   Y_{T^c\cap \mathcal{G}^{h-1}}, \{ Y_{\{\ell\}\times T^c\cap \mathcal{G}_h}\}_{\ell<i}, E_{\mathcal{I}\times \mathcal{I}}) \notag\\
& \le  H_{p_{ X_{\mathcal{I}},  Y_{\mathcal{I}}}}(Y_{\{i\}\times (T^c\cap \mathcal{G}_h)}| Y_{\{i\}\times T^c\cap \mathcal{G}^{h-1}}, E_{\mathcal{I}\times \mathcal{I}}) \notag\\
 &\qquad -H_{p_{ X_{\mathcal{I}},  Y_{\mathcal{I}}}}(Y_{\{i\}\times (T^c\cap \mathcal{G}_h)}|  X_{\mathcal{S}^h},   Y_{\mathcal{G}^{h-1}}, \{ Y_{\{\ell\}\times T^c\cap \mathcal{G}_h}\}_{\ell<i}, E_{\mathcal{I}\times \mathcal{I}}) \notag\\
&\stackrel{\eqref{markovChainWENNetwork}}{=} H_{p_{ X_{\mathcal{I}},  Y_{\mathcal{I}}}}(Y_{\{i\}\times (T^c\cap \mathcal{G}_h)}|  Y_{\{i\}\times T^c\cap \mathcal{G}^{h-1}}, E_{\mathcal{I}\times \mathcal{I}}) - H_{p_{ X_{\mathcal{I}},  Y_{\mathcal{I}}}}(Y_{\{i\}\times (T^c\cap \mathcal{G}_h)}|X_{i}, E_{\mathcal{I}\times \mathcal{I}})\notag\\
& \le I_{p_{ X_{\mathcal{I}},  Y_{\mathcal{I}}}}(X_{i};Y_{\{i\}\times (T^c\cap \mathcal{G}_h)}| Y_{\{i\}\times T^c\cap \mathcal{G}^{h-1}}, E_{\mathcal{I}\times \mathcal{I}}).
\end{align*}
\item[(c)] follows from \eqref{PrYijGivenXij} and \eqref{defEij} that $Y_{\{i\}\times (T^c\cap \mathcal{G}_h)}$ is a function of $(X_i, E_{\mathcal{I}\times \mathcal{I}})$.
\end{enumerate}
Following \eqref{eqn4WENCon} and letting $\mathbf{1}^{T^c}$ denote the $|T^c|$-dimensional all-1 tuple, we consider the following chain of inequalities for each $h\in\{1, 2, \ldots, \alpha\}$ and each $i\in T\cap\mathcal{S}^h$:
\begin{align}
&H_{p_{ X_{\mathcal{I}},  Y_{\mathcal{I}}}}(Y_{\{i\}\times T^c}| E_{\{i\}\times T^c}) \notag\\
& = \Pr\{E_{\{i\} \times T^c} =\mathbf{1}^{T^c} \}H_{p_{X_\mathcal{I}, Y_\mathcal{I}}}(Y_{\{i\}\times T^c} |E_{\{i\} \times T^c}=\mathbf{1}^{T^c}) \notag\\
 & \qquad + \Pr\{E_{\{i\} \times T^c} \ne \mathbf{1}^{T^c} \}H_{p_{X_\mathcal{I}, Y_\mathcal{I}}}(Y_{\{i\}\times T^c} |E_{\{i\} \times T^c}\ne\mathbf{1}^{T^c})\notag\\
& \stackrel{\text{(a)}}{=} \Pr\{E_{\{i\} \times T^c} \ne \mathbf{1}^{T^c} \}H_{p_{X_\mathcal{I}, Y_\mathcal{I}}}(X_i |E_{\{i\} \times T^c}\ne\mathbf{1}^{T^c})\notag\\
& \le \Pr\{E_{\{i\} \times T^c} \ne \mathbf{1}^{T^c} \}|\mathcal{X}_i|\notag\\
&\stackrel{\text{(b)}}{=} \left(1-\prod_{j\in T^c}e_{i,j} \right)|\mathcal{X}_i|  \label{eqnWENconditionalH=0eq5Con}
\end{align}
where
\begin{enumerate}
\item[(a)] follows from \eqref{defEij} that for each $j\in \mathcal{I}$,
\[
Y_{i,j}=
\begin{cases}
\varepsilon & \text{if $E_{i,j}=1$,} \\
X_i & \text{otherwise.}
\end{cases}
\]
        \item[(b)] follows from \eqref{EIIindependetOfXI*} and \eqref{EIIindependetOfXI}.
\end{enumerate}
Combining \eqref{eqn4WENCon} and \eqref{eqnWENconditionalH=0eq5Con}, we have
\begin{align}
 & \sum_{h=1}^\alpha  I_{p_{ X_{\mathcal{I}},  Y_{\mathcal{I}}}}( X_{T\cap \mathcal{S}^h},  Y_{T\cap \mathcal{G}^{h-1}} ;   Y_{T^c}|  X_{T^c \cap \mathcal{S}^h},   Y_{T^c\cap \mathcal{G}^{h-1}}, E_{\mathcal{I}\times \mathcal{I}}) \notag\\
 & \le \sum_{i\in T\cap \mathcal{S}^h} \left(1-\prod_{j\in T^c}e_{i,j} \right)|\mathcal{X}_i| \notag\\
 & \le  \sum_{i\in T}  \left(1-\prod_{j\in T^c}e_{i,j} \right)|\mathcal{X}_i|. \label{eqnWENconditionalH=0eq6Con}
\end{align}

Consequently, it follows from \eqref{defRWEN}, \eqref{RoutWEN} and \eqref{eqnWENconditionalH=0eq6Con} that \eqref{RoutSubsetRinWEN} holds, which implies from \eqref{converseWENByTheorem2} that $\mathcal{C}\subseteq \mathcal{R}_{\text{in}}^{\text{WEN}} $.


\section{Concluding Remarks} \label{conclusion}
We have investigated under the generalized-delay model three classes of delay-independent multimessage multicast networks (MMNs), namely the deterministic MMN dominated by product distributions, the MMN consisting of independent DMCs and the wireless erasure network respectively. We are able to evaluate the capacity regions for the above classes of MMNs with zero-delay nodes and demonstrate that their capacity regions coincide with the positive-delay regions, which implies that the above classes of MMNs with zero-delay nodes belong to the category of delay-independent MMNs. In other words, for each MMN with zero-delay nodes which belongs to one of the above three classes, the set of achievable rate tuples does not depend on the delay amounts incurred by the nodes in the network. This is in contrast to the fact that for some MMNs with zero-delay nodes, the set of achievable rate tuples shrinks if we impose the additional constraint that each node incurs a positive delay. An important implication of our result is that for each MMN belonging to one of the above three classes, using different methods for handling delay and synchronization does not affect the network capacity.

Future research may continue the theme of this work -- to identify other important classes of delay-independent and delay-dependent MMNs under the generalized-delay model. This work is limited to identifying delay-independent MMNs whose capacity regions lie in the corresponding cut-set bounds and at the same time the cut-set bounds can be achieved. The search of delay-independent and delay-dependent MMNs whose capacity regions are strictly smaller than the classical cut-set bounds is an interesting research direction. Another direction is exploring delay-dependent MMNs whose capacity regions are strictly larger than the classical cut-set bounds.

\appendices

\section{Proof of Lemma~\ref{cutsetMCLemma*}}\label{appendixA}
Fix a $(B, n, M_{\mathcal{I}\times\mathcal{I}})$-code, and let $p_{W_{\mathcal{I}},X_{\mathcal{I}}^n, Y_{\mathcal{I}}^n  }$ be the joint distribution induced by the code according to Definitions~\ref{defCode} and~\ref{defDiscreteMemoryless}. For each $k\in\{1, 2, \ldots, n\}$, let $U^{k-1}\triangleq (W_{\mathcal{I}}, X_{\mathcal{I}}^{k-1}, Y_{\mathcal{I}}^{k-1})$ be the collection of random variables that are generated before the $k^{\text{th}}$ time slot for the $(B, n, M_{\mathcal{I}\times\mathcal{I}})$-code. In order to prove \eqref{lemmaStatement1MCLemma*}, it suffices to show that
\begin{equation}
H_{p_{U^{k-1},X_{\mathcal{S}^h,k}, Y_{\mathcal{G}^h,k} }}(X_{\mathcal{S}^h,k}, Y_{\mathcal{G}^h,k} |  U^{k-1})=0
\label{cutsetstatement2***}
\end{equation}
holds for each $k\in\{1, 2, \ldots, n\}$ and each $h\in\{1, 2, \ldots, \alpha\}$, which will then imply that
\begin{align*}
H_{p_{W_{\mathcal{I}}, X_{\mathcal{I}}^n, Y_{\mathcal{I}}^n}}(X_{\mathcal{I}}^n, Y_{\mathcal{I}}^n |  W_{\mathcal{I}}) &=   \sum_{k=1}^n H_{p_{U^{k-1}, X_{\mathcal{I},k}, Y_{\mathcal{I},k}}}(X_{\mathcal{I},k}, Y_{\mathcal{I},k}|U^{k-1}) \notag\\
&\stackrel{\eqref{cutsetstatement2***}}{=}0.
\end{align*}
Fix a $k\in\{1, 2, \ldots, n\}$. We prove \eqref{cutsetstatement2***} by induction on~$h$ as follows. For $h=1$, the LHS of \eqref{cutsetstatement2***} is
\begin{align}
&H_{p_{U^{k-1},X_{\mathcal{S}^1,k}, Y_{\mathcal{G}^1,k} }}(X_{\mathcal{S}^1,k}, Y_{\mathcal{G}^1,k} |  U^{k-1}) \notag\\
& = H_{p_{U^{k-1},X_{\mathcal{S}^1,k}}}(X_{\mathcal{S}^1,k}|  U^{k-1}) + H_{p_{U^{k-1},X_{\mathcal{S}^1,k},Y_{\mathcal{G}^1,k}}}(Y_{\mathcal{G}^1,k}|U^{k-1},X_{\mathcal{S}^1,k})\notag\\
& \stackrel{\text{(a)}}{=}  H_{p_{U^{k-1},X_{\mathcal{S}^1,k},Y_{\mathcal{G}^1,k}}}(Y_{\mathcal{G}^1,k}|U^{k-1},X_{\mathcal{S}^1,k})\notag \\
& \le H_{p_{X_{\mathcal{S}^1,k}, Y_{\mathcal{G}^1,k}}}(Y_{\mathcal{G}^1,k}|X_{\mathcal{S}^1,k}) \notag\\
& \stackrel{\eqref{memorylessStatement}}{=} H_{p_{X_{\mathcal{S}^1,k}}q_{Y_{\mathcal{G}^1,k}|X_{\mathcal{S}^1,k}}^{(1)}}(Y_{\mathcal{G}^1,k}|X_{\mathcal{S}^1,k}) \notag\\
& \stackrel{\text{(b)}}{=}  0,\label{MIstatement1}
\end{align}
where
\begin{enumerate}
\item[(a)] follows from Definitions~\ref{defFeasible} and~\ref{defCode} that $X_{\mathcal{S}^1,k}$ is a function of $U^{k-1}$ for the code.
    \item[(b)] follows from the fact that $q^{(1)}$ is deterministic (cf.\ Definition~\ref{definitionDeterministicMMN}).
    \end{enumerate}
If \eqref{cutsetstatement2***} holds for $h=m$, i.e.,
\begin{equation}
H_{p_{U^{k-1},X_{\mathcal{S}^m,k}, Y_{\mathcal{G}^m,k}}}(X_{\mathcal{S}^m,k}, Y_{\mathcal{G}^m,k} |  U^{k-1})=0 \label{assumption2***}
\end{equation}
then for $h=m+1$ such that $m+1\le \alpha$, the LHS of \eqref{cutsetstatement2***} is
\begin{align}
&H_{p_{U^{k-1},X_{\mathcal{S}^{m+1},k}, Y_{\mathcal{G}^{m+1},k}}}(X_{\mathcal{S}^{m+1},k}, Y_{\mathcal{G}^{m+1},k} |  U^{k-1})\notag \\
& = H_{p_{U^{k-1},X_{\mathcal{S}^m,k},Y_{\mathcal{G}^{m},k}}}(X_{\mathcal{S}^m,k},Y_{\mathcal{G}^{m},k} |U^{k-1} )+ H_{p_{U^{k-1},X_{\mathcal{S}^m,k},Y_{\mathcal{G}^{m},k} ,X_{\mathcal{S}_{m+1},k}}}(X_{\mathcal{S}_{m+1},k}|U^{k-1},X_{\mathcal{S}^m,k},Y_{\mathcal{G}^{m},k} ) \notag\\
&\qquad + H_{p_{U^{k-1},X_{\mathcal{S}^{m+1},k},Y_{\mathcal{G}^{m+1},k}}}(Y_{\mathcal{G}_{m+1},k}|U^{k-1},X_{\mathcal{S}^{m+1},k}, Y_{\mathcal{G}^{m},k}) \notag\\
& \stackrel{\eqref{assumption2***}}{=}  H_{p_{U^{k-1},X_{\mathcal{S}^m,k},Y_{\mathcal{G}^{m},k} ,X_{\mathcal{S}_{m+1},k}}}(X_{\mathcal{S}_{m+1},k}|U^{k-1},X_{\mathcal{S}^m,k},Y_{\mathcal{G}^{m},k} ) \notag\\*
&\qquad + H_{p_{U^{k-1},X_{\mathcal{S}^{m+1},k},Y_{\mathcal{G}^{m+1},k}}}(Y_{\mathcal{G}_{m+1},k}|U^{k-1},X_{\mathcal{S}^{m+1},k}, Y_{\mathcal{G}^{m},k})\notag\\
& \stackrel{\text{(a)}}{=}  H_{p_{U^{k-1},X_{\mathcal{S}^{m+1},k},Y_{\mathcal{G}^{m+1},k}}}(Y_{\mathcal{G}_{m+1},k}|U^{k-1},X_{\mathcal{S}^{m+1},k}, Y_{\mathcal{G}^{m},k})\notag \\
& \le H_{p_{X_{\mathcal{S}^{m+1},k},Y_{\mathcal{G}^{m+1},k}}}(Y_{\mathcal{G}_{m+1},k}|X_{\mathcal{S}^{m+1},k}, Y_{\mathcal{G}^{m},k}) \notag\\
&\stackrel{\eqref{memorylessStatement}}{=}  H_{p_{X_{\mathcal{S}^{m+1},k},Y_{\mathcal{G}_{m},k}}q_{Y_{\mathcal{G}_{m+1},k}|X_{\mathcal{S}^{m+1},k},Y_{\mathcal{G}_{m},k}}^{(m+1)}}(Y_{\mathcal{G}_{m+1},k}|X_{\mathcal{S}^{m+1},k}, Y_{\mathcal{G}^{m},k})\notag\\
& \stackrel{\text{(b)}}{=}  0, \label{MIstatement2}
\end{align}
where
\begin{enumerate}
\item[(a)] follows from Definitions~\ref{defFeasible} and~\ref{defCode} that $X_{\mathcal{S}_{m+1},k}$ is a function of $(U^{k-1}, Y_{\mathcal{G}^{m},k})$ for the code.
    \item[(b)] follows from the fact that $q^{(m+1)}$ is deterministic (cf.\ Definition~\ref{definitionDeterministicMMN}).
\end{enumerate}
   For $h=1$, it follows from \eqref{MIstatement1} that \eqref{cutsetstatement2***} holds.  For all $1\le m \le \alpha-1$, it follows from \eqref{assumption2***} and \eqref{MIstatement2} that if \eqref{cutsetstatement2***} is assumed to be true for $h=m$, then \eqref{cutsetstatement2***} is also true for $h=m+1$. Consequently, it follows by  mathematical induction that \eqref{cutsetstatement2***} holds for all $1\le h\le \alpha$.

\section{Proof of the Achievability of Theorem~\ref{thmWEN}}\label{appendixB}
Our goal is to prove \eqref{achievabilityStatementWEN}. Let $u_{X_i}$ be the uniform distribution on $\mathcal{X}_i$ for each $i\in\mathcal{I}$ and let
   \begin{equation}
   u_{X_\mathcal{I}, Y_\mathcal{I}} = \left(\prod_{i=1}^N u_{X_i}\right)\left(\prod_{h=1}^\alpha q_{Y_{\mathcal{G}_h}|X_\mathcal{S}^{h}, Y_\mathcal{G}^{h-1}}^{(h)}\right). \label{productDistributionWEN}
   \end{equation}
 Fix any $T\subseteq\mathcal{I}$ such that
   \begin{equation}
   T^c\cap \mathcal{D}\ne \emptyset. \label{eqnWENTCintersectD}
   \end{equation}
In order to apply Theorem~\ref{theoremNNC}, we consider
   \begin{align}
  & H_{u_{X_\mathcal{I}, Y_\mathcal{I}}}(Y_T|X_\mathcal{I},Y_{T^c}) \notag\\
  & \stackrel{\text{(a)}}{=} H_{u_{X_\mathcal{I}, Y_\mathcal{I}}}(Y_T|X_\mathcal{I},Y_{T^c}, E_{\mathcal{I}\times\mathcal{I}}) \notag\\
  & \stackrel{\text{(b)}}{=} 0 \label{eqnWENconditionalH=0eq1}
   \end{align}
   and
    \begin{align}
  & I_{u_{X_\mathcal{I}, Y_\mathcal{I}}}(X_T;Y_{T^c}|X_{T^c}) \notag\\*
  & \stackrel{\text{(c)}}{=} I_{u_{X_\mathcal{I}, Y_\mathcal{I}}}(X_T;Y_{T^c},E_{\mathcal{I}\times\mathcal{I}} |X_{T^c}) \notag\\
  & \stackrel{\text{(d)}}{=} I_{u_{X_\mathcal{I}, Y_\mathcal{I}}}(X_T;Y_{T^c} |X_{T^c},E_{\mathcal{I}\times\mathcal{I}}) \notag\\
   & = H_{u_{X_\mathcal{I}, Y_\mathcal{I}}}(Y_{T^c} |X_{T^c},E_{\mathcal{I}\times\mathcal{I}}) - H_{u_{X_\mathcal{I}, Y_\mathcal{I}}}(Y_{T^c} |X_{\mathcal{I}},E_{\mathcal{I}\times\mathcal{I}}) \notag\\
   & \stackrel{\eqref{eqnWENconditionalH=0eq1}}{=}  H_{u_{X_\mathcal{I}, Y_\mathcal{I}}}(Y_{T^c} |X_{T^c},E_{\mathcal{I}\times\mathcal{I}}) \notag\\
   & \stackrel{\eqref{defWENalphabetYm}}{=} H_{u_{X_\mathcal{I}, Y_\mathcal{I}}}(\{Y_{i,j}\}_{(i,j)\in \mathcal{I}\times T^c} |X_{T^c},E_{\mathcal{I}\times\mathcal{I}})  \notag\\
   &\stackrel{\text{(e)}}{=}  H_{u_{X_\mathcal{I}, Y_\mathcal{I}}}(\{Y_{i,j}\}_{(i,j)\in T\times T^c} |X_{T^c},E_{\mathcal{I}\times\mathcal{I}}) \label{eqnWENconditionalH=0eq2}
   \end{align}
   where
   \begin{enumerate}
   \item[(a)] follows from Statements~(i) and~(ii) in the previous subsection and \eqref{eqnWENTCintersectD} that $Y_{T^c}$ contains the random variable $E_{\mathcal{I}\times\mathcal{I}}$.
\item[(b)] follows from \eqref{PrYijGivenXij} and \eqref{defEij} that $Y_T$ is a function of $(X_\mathcal{I}, E_{\mathcal{I}\times\mathcal{I}})$.
    \item[(c)] follows from Statements~(i) and~(ii) in the previous subsection and \eqref{eqnWENTCintersectD} that $Y_{T^c}$ contains the random variable $E_{\mathcal{I}\times\mathcal{I}}$.
        \item[(d)] follows from \eqref{EIIindependetOfXI} that $X_\mathcal{I}$ and $E_{\mathcal{I}\times\mathcal{I}}$ are independent, i.e.,
            \begin{equation}
            I_{u_{X_\mathcal{I}, Y_\mathcal{I}}}(X_\mathcal{I};E_{\mathcal{I}\times\mathcal{I}})=0. \label{XEindependentDistU}
            \end{equation}
             \item[(e)] follows from \eqref{PrYijGivenXij} and \eqref{defEij} that $\{Y_{i,j}\}_{(i,j)\in T^c\times T^c}$ is a function of $(X_{T^c}, E_{\mathcal{I}\times\mathcal{I}})$.
\end{enumerate}
In order to further simplify \eqref{eqnWENconditionalH=0eq2}, consider the following chain of inequalities for any $T_1, T_2\subseteq\mathcal{I}$ such that $T_1\cap T_2 = \emptyset$:
\begin{align}
    &I_{u_{X_\mathcal{I}, Y_\mathcal{I}}}(\{Y_{i,j}\}_{(i,j)\in T_1\times T_2}; X_{T_2}|E_{\mathcal{I}\times \mathcal{I}}) \notag\\
   & \le I_{u_{X_\mathcal{I}, Y_\mathcal{I}}}(X_{T_1}, \{Y_{i,j}\}_{(i,j)\in T_1\times T_2}; X_{T_2}|E_{\mathcal{I}\times \mathcal{I}}) \notag\\
    & \stackrel{\text{(a)}}{=} I_{u_{X_\mathcal{I}, Y_\mathcal{I}}}(X_{T_1}; X_{T_2}|E_{\mathcal{I}\times \mathcal{I}})  \notag\\
    &\stackrel{\text{(b)}}{=} I_{u_{X_\mathcal{I}, Y_\mathcal{I}}}(X_{T_1}; X_{T_2})  \notag\\
    & \stackrel{\text{(c)}}{=} 0
  \label{eqnWENconditionalH=0eq3}
\end{align}
where
\begin{enumerate}
\item[(a)] follows from the fact that $\{Y_{i,j}\}_{(i,j)\in T_1\times T_2}$ is a function of $(X_{T_1}, E_{\mathcal{I}\times \mathcal{I}})$.
\item[(b)] follows from \eqref{EIIindependetOfXI} that $X_\mathcal{I}$ and $E_{\mathcal{I}\times \mathcal{I}}$ are independent.
    \item[(c)] follows from \eqref{productDistributionWEN} that $X_{T_1}$ and $X_{T_2}$ are independent.
\end{enumerate}
Following \eqref{eqnWENconditionalH=0eq2}, consider the following chain of inequalities:
 \begin{align}
     &H_{u_{X_\mathcal{I}, Y_\mathcal{I}}}(\{Y_{i,j}\}_{(i,j)\in T\times T^c} |X_{T^c},E_{\mathcal{I}\times\mathcal{I}}) \notag\\
     &\stackrel{\text{(a)}}{=}H_{u_{X_\mathcal{I}, Y_\mathcal{I}}}(\{Y_{i,j}\}_{(i,j)\in T\times T^c} |X_{T^c},E_{\mathcal{I}\times\mathcal{I}}) + I_{u_{X_\mathcal{I}, Y_\mathcal{I}}}(X_{T^c};\{Y_{i,j}\}_{(i,j)\in T\times T^c} |E_{\mathcal{I}\times\mathcal{I}}) \notag\\
     & = H_{u_{X_\mathcal{I}, Y_\mathcal{I}}}(\{Y_{i,j}\}_{(i,j)\in T\times T^c} |E_{\mathcal{I}\times\mathcal{I}})  \notag\\
   & = \sum_{i\in T} H_{u_{X_\mathcal{I}, Y_\mathcal{I}}}(Y_{\{i\}\times T^c} |E_{\mathcal{I}\times\mathcal{I}}, \{Y_{m,\ell}\}_{m\in T, m<i, \ell\in T^c}) \notag\\
      & \ge \sum_{i\in T} H_{u_{X_\mathcal{I}, Y_\mathcal{I}}}(Y_{\{i\}\times T^c} |E_{\mathcal{I}\times\mathcal{I}}, \{Y_{m,\ell}\}_{m\in T, m<i, \ell\in T^c}, \{X_m\}_{m\in T, m<i}) \notag\\
   & \stackrel{\text{(b)}}{=}  \sum_{i\in T} H_{u_{X_\mathcal{I}, Y_\mathcal{I}}}(Y_{\{i\}\times T^c} |E_{\mathcal{I}\times\mathcal{I}}, \{X_m\}_{m\in T, m<i})\notag\\
   &\stackrel{\eqref{eqnWENconditionalH=0eq3}}{=}\sum_{i\in T} H_{u_{X_\mathcal{I}, Y_\mathcal{I}}}(Y_{\{i\}\times T^c} |E_{\mathcal{I}\times\mathcal{I}})\notag\\
   & \stackrel{\text{(c)}}{\ge}  \sum_{i\in T} (H_{u_{X_\mathcal{I}, Y_\mathcal{I}}}(Y_{\{i\}\times T^c} |E_{\{i\}\times T^c}) - I_{u_{X_\mathcal{I}, Y_\mathcal{I}}}(E_{\{i\}\times T^c};Y_{\{i\}\times T^c} |X_{i}, E_{\{i\}\times T^c}))\notag\\
   &\stackrel{\text{(d)}}{=}H_{u_{X_\mathcal{I}, Y_\mathcal{I}}}(Y_{\{i\}\times T^c} |E_{\{i\}\times T^c})
   \label{eqnWENconditionalH=0eq4}
   \end{align}
   where
   \begin{enumerate}
   \item[(a)] follows from \eqref{eqnWENconditionalH=0eq3} by letting $T_1 = T$ and $T_2=T^c$.
            \item[(b)] follows from \eqref{PrYijGivenXij} and \eqref{defEij} that $\{Y_{m,\ell}\}_{m\in T, m<i, \ell\in T^c}$ is a function of $(\{X_m\}_{m\in T, m<i}, E_{\mathcal{I}\times\mathcal{I}})$ .
                \item[(c)] follows from the fact that
                \begin{align*}
                &I_{u_{X_\mathcal{I}, Y_\mathcal{I}}}(E_{\{i\}\times T^c};Y_{\{i\}\times T^c} |E_{\{i\}\times T^c}) \notag\\
                & \le I_{u_{X_\mathcal{I}, Y_\mathcal{I}}}(E_{\{i\}\times T^c};X_{i}, Y_{\{i\}\times T^c} |E_{\{i\}\times T^c}) \notag\\
                & \stackrel{\eqref{XEindependentDistU}}{=}  I_{u_{X_\mathcal{I}, Y_\mathcal{I}}}(E_{\{i\}\times T^c};Y_{\{i\}\times T^c} |X_{i}, E_{\{i\}\times T^c}).
                \end{align*}
                \item[(d)] follows from  \eqref{PrYijGivenXij} and \eqref{defEij} that $Y_{\{i\}\times T^c}$ is a function of $(X_{i}, E_{\{i\}\times T^c})$.
\end{enumerate}
Following \eqref{eqnWENconditionalH=0eq4} and letting $\mathbf{1}^{T^c}$ denote the $|T^c|$-dimensional all-1 tuple, we consider the following chain of equalities for each $i\in T$:
\begin{align}
&H_{u_{X_\mathcal{I}, Y_\mathcal{I}}}(Y_{\{i\}\times T^c} |E_{\{i\}\times T^c}) \notag\\
& = \Pr\{E_{\{i\}\times T^c} =\mathbf{1}^{T^c} \}H_{u_{X_\mathcal{I}, Y_\mathcal{I}}}(Y_{\{i\}\times T^c} |E_{\{i\}\times T^c}=\mathbf{1}^{T^c}) \notag\\
 & \qquad + \Pr\{E_{\{i\}\times T^c} \ne \mathbf{1}^{T^c} \}H_{u_{X_\mathcal{I}, Y_\mathcal{I}}}(Y_{\{i\}\times T^c} |E_{\{i\}\times T^c}\ne\mathbf{1}^{T^c})\notag\\
& \stackrel{\text{(a)}}{=} \Pr\{E_{\{i\}\times T^c} \ne \mathbf{1}^{T^c} \}H_{u_{X_\mathcal{I}, Y_\mathcal{I}}}(X_i |E_{\{i\}\times T^c}\ne\mathbf{1}^{T^c})\notag\\
& \stackrel{\text{(b)}}{=} \Pr\{E_{\{i\}\times T^c} \ne \mathbf{1}^{T^c} \}H_{u_{X_\mathcal{I}, Y_\mathcal{I}}}(X_i)\notag\\
& \stackrel{\text{(c)}}{=} \Pr\{E_{\{i\}\times T^c} \ne \mathbf{1}^{T^c} \}|\mathcal{X}_i|\notag\\
&\stackrel{\text{(d)}}{=} \left(1-\prod_{j\in T^c}e_{i,j} \right)|\mathcal{X}_i|  \label{eqnWENconditionalH=0eq5}
\end{align}
where
\begin{enumerate}
\item[(a)] follows from \eqref{defEij} that for each $j\in \mathcal{I}$,
\[
Y_{i,j}=
\begin{cases}
\varepsilon & \text{if $E_{i,j}=1$,} \\
X_i & \text{otherwise.}
\end{cases}
\]
\item[(b)] follows from \eqref{EIIindependetOfXI} that $X_\mathcal{I}$ and $E_{\mathcal{I}\times \mathcal{I}}$ are independent.
    \item[(c)] follows from \eqref{productDistributionWEN} that $X_i$ is uniform on $|\mathcal{X}_i|$.
        \item[(d)] follows from \eqref{EIIindependetOfXI*} and \eqref{EIIindependetOfXI}.
\end{enumerate}
Combining \eqref{eqnWENconditionalH=0eq2}, \eqref{eqnWENconditionalH=0eq4} and  \eqref{eqnWENconditionalH=0eq5}, we have
\begin{equation}
 I_{u_{X_\mathcal{I}, Y_\mathcal{I}}}(X_T;Y_{T^c}|X_{T^c}) \ge \sum_{i\in T}  \left(1-\prod_{j\in T^c}e_{i,j} \right)|\mathcal{X}_i|. \label{eqnWENconditionalH=0eq6}
\end{equation}
Using Theorem~\ref{theoremNNC}, \eqref{defRWEN}, \eqref{eqnWENconditionalH=0eq1} and \eqref{eqnWENconditionalH=0eq6},  we have $\mathcal{R}_{\text{in}}^{\text{WEN}} \subseteq \mathcal{C}_+$.


\section*{Acknowledgment}
The author would also like to thank Associate Editor Sae-Young Chung and the two anonymous reviewers for their useful comments that improve the
presentation of this work.

\ifCLASSOPTIONcaptionsoff
  \newpage
\fi

\end{document}